\documentclass[usenatbib,letter]{mnras}
\usepackage{graphicx, amssymb, aas_macros, natbib, bm}

\usepackage{mathptmx}
\usepackage{epsfig}
\usepackage{wrapfig}
\usepackage{amsmath}
\usepackage{natbib}
\usepackage{graphicx}
\usepackage{subfigure}
\usepackage{enumitem}

\newcommand\msun{\hbox{\ensuremath{{\rm M}_{\odot}}}}

\def\msa{\hbox{\ensuremath{\rm mag~arcsec^{-2}}}}

\newcommand{\scinot}[1]{\ensuremath{\times 10^{#1}}}

\newcommand{\cse}{Computing in Science \& Engineering}

 \title[Tidally-Stripped UDGs]{The Formation of Ultra Diffuse Galaxies in
   Cored Dark Matter 
 Halos Through Tidal Stripping and Heating}
 \author[Carleton et al.]
 {Timothy Carleton,$^{1,3}$\thanks{$\!\!$e-mail: carletont@missouri.edu}
 	Rapha\"el Errani,$^{2}$
 	\newauthor 	Michael Cooper,$^{1}$
 	Manoj Kaplinghat,$^{1}$
 	 Jorge Pe\~narrubia,$^{2}$
 	 Yicheng Guo$^{3}$\\ \\
 	$\!\!^1$Center for Cosmology, Department of Physics and Astronomy, 
 	 4129 Reines Hall, University of California, Irvine, CA 92697,
 	 USA \\
 	$\!\!^{2}$Institute for Astronomy, University of Edinburgh, Royal Observatory, Blackford Hill, Edinburgh EH9 3HJ, UK \\	
 	$\!\!^{3}$Department of Physics and Astronomy, 223 Physics Building, University of Missouri, Columbia, MO 65211, USA \\	
 }
 
\begin{document}
 	
 	\pagerange{\pageref{firstpage}--\pageref{lastpage}} 
 	\pubyear{2018}
 	
 	\maketitle

 	\begin{abstract}
          We propose that the Ultra-Diffuse Galaxy (UDG) population represents a
          set of satellite galaxies born in $\sim10^{10}-10^{11}$~\msun{}
          halos, similar to field dwarfs, which suffer a dramatic reduction in
          surface brightness due to tidal stripping and heating. This scenario is
          observationally motivated by the radial alignment of UDGs in Coma
          as well as the significant dependence of UDG abundance on cluster
          mass. As a test of this formation scenario, we apply a semi-analytic
          model describing the change in stellar mass and half-light radius of
          dwarf satellites, occupying either cored or cuspy halos, to cluster
          subhalos in the Illustris-dark simulation. Key to this model are results from
          simulations which indicate that galaxies in cored dark-matter halos
          expand significantly in response to tidal stripping and heating, whereas galaxies
          in cuspy halos experience limited size evolution. Our analysis
          indicates that a population of tidally-stripped dwarf galaxies,
          residing in cored halos (like those hosting low-surface brightness field
          dwarfs), is able to reproduce the observed sizes and stellar
          masses of UDGs in clusters remarkably well.
 	\end{abstract}
 	
 	
 	\section{Introduction}
 	\label{sec:intro}
        The recent identification of a large population of ultra-diffuse
        galaxies (UDGs) in the Coma cluster \citep{vandokkum2015,vandokkum2015b}
        has led to a renewed interest in low surface brightness galaxies. Since
        the original identification of $47$ UDGs in Coma, hundreds more have
        been identified in the Coma, Perseus, Fornax, and Virgo clusters
        \citep{koda2015, mihos2015, munoz2015, yagi2016, 
          wittmann2017,venhola2017,ordenes-briceno2018,eigenthaler2018} as well as in higher-redshift clusters 
        \citep{lee2017, janssens2017} and lower-mass
        groups \citep{martin2016, merritt2016, smith2016, toloba2016,
          roman2017iso, shi2017, spekkens2018,muller2018}. These cluster UDGs are characterized by old
        stellar populations, stellar masses of $10^7-10^8~\msun$, and
        exponential light profiles similar to typical low-mass dwarf
        ellipticals, but with half-light radii $>1.5$~kpc and central surface
        brightnesses $>24$~\msa{}. Although UDGs represent a minority of the
        overall dwarf galaxy population, the peculiar nature of these systems
        provides an excellent testbed for our understanding of galaxy evolution.

        The unusual nature of these systems has sparked a number of theories
        regarding their formation. In particular, the dark matter content of
        UDGs has been subject to extensive debate, with some investigations,
        using both stellar velocity dispersions \citep{vandokkum2016} and H{\sc
          i} dynamics \citep{trujillo2017}, arriving at the conclusion that UDGs
        are extremely dark matter dominated in their central regions, in
        agreement with globular cluster-based mass determinations
        \citep{peng2016,beasley2016,vandokkum2017,toloba2018}.  Other studies, also
        using globular cluster velocity dispersions \citep{vandokkum2018}, as well
        as globular cluster colors and abundances \citep{amorisco2016b,
          beasley2016b,amorisco2018c} have found that the mass-to-light ratio of UDGs is
        consistent with that of similarly massive dwarfs (however, \cite{laporte2018} and \cite{martin2018} have cautioned against the use a limited number of globular clusters to estimate the dynamical mass of UDGs, citing significant systematic uncertainties). This latter conclusion is consistent with weak
        gravitational lensing limits \citep{sifon2017} and observations of the
        distribution of UDGs in groups and clusters \citep{roman2017}. Additional evidence that UDGs live in 
        dwarf-scale halos derives from the abundance of UDGs in clusters: there simply are not enough 
        Milky~Way-mass halos in clusters to host $90\%$ of the UDG population \citep{amorisco2018}.

        Prompted by the observation of a large internal velocity dispersion for
        a UDG in Coma, \cite{vandokkum2016} hypothesized a formation scenario in
        which UDGs are born within Milky~Way-like dark matter halos but fail to
        form most of their stars due to extreme feedback \citep[see
        also][]{agertz2016}. Similarly, it has been suggested that UDGs are formed when strong feedback expands 
        dwarf galaxies, which retain their large sizes as they fall into a cluster \citep{dicintio2017,chan2017}. The 
        feedback required in this scenario finds
        support in recent simulations in which dwarf galaxies experience cycles
        of expansion and contraction due to large stellar feedback effects
        \citep{pontzen2014}. These simulations with extreme feedback, however,
        overpredict the sizes of isolated dwarf galaxies \citep{lange2015} and
        are not able to reproduce the relative abundance of cluster vs. field UDGs. Alternatively, it
        has been suggested that UDGs represent galaxies living in halos in the
        tail of the spin distribution \citep{amorisco2016,leisman2017}. While
        this scenario does a better job of reproducing the observed size
        distribution, it predicts that cluster UDGs should be less abundant and biased towards
        less-diffuse systems compared with UDGs in the field, contrary to
        observations \citep{roman2017iso,jones2018}.
 		
        Clues to the UDG formation mechanism have come as more complete
        observations of the environments hosting UDGs have been
        conducted. Although UDGs are observed outside cluster environments
        \citep{leisman2017,roman2017iso,williams2016}, cluster UDGs are both more diffuse and
        relatively more abundant than isolated UDGs. A systematic survey of UDGs
        in clusters from \cite{vanderburg2017} found that the relative abundance
        of UDGs increases with increasing cluster mass, consistent with an
        increase in the mass fraction of UDGs from $\sim0.1\%$ in the field
        \citep{leisman2017,jones2018} to $\sim3\%$ in clusters
        \citep{vanderburg2016,yagi2016}. Additionally, \cite{roman2017iso} found
        that UDGs within $300$~kpc of the Hickson Compact Groups HCG07, HCG25,
        and HCG98 have lower central surface brightnesses compared with UDGs at
        farther distances from the group center. Furthermore, there is a
        significant radial alignment observed in Coma UDGs
        (\citealt{yagi2016}; but see \citealt{venhola2017,eigenthaler2018}), suggesting that the shapes of these
        systems have been affected by tidal interactions. Lastly,
        \cite{burkert2017} show that the axis ratios of Coma UDGs are more
        consistent with elongated structures than puffed-up disks. These
        distinctions strongly suggest that environmental mechanisms are at play
        in the formation and evolution of UDGs. In fact, `galaxy harassment,' tidal stripping, and ram-pressure 
        stripping have recently been suggested as possible UDG formation mechanisms 
        \citep{safarzadeh2017,conselice2018,ogiya2018,bennet2018}.
 		
        Within the Local Group, studies have shown that satellite galaxies are
        significantly influenced by the tidal effects of a massive host
        galaxy. For example, \cite{penarrubia2010} find that satellites orbiting
        the Milky~Way and Andromeda can lose over $99\%$ of their mass during
        the course of a few orbits due to tidal stripping. Importantly, they
        show that tidal stripping affects cored and cuspy halos differently:
        cored halos endure a greater degree of stripping than cuspy halos, and
        cuspy halos experience stripping primarily in the halo outskirts,
        whereas tidal stripping of cored halos occurs more evenly throughout the halo.
        Following this work, \cite{errani2015} investigated the effect of tidal
        stripping on satellite galaxies. While the stellar population of
        satellites in both cored and cuspy halos experiences expansion as a
        consequence of tidal heating, as well as stellar stripping, satellites
        in cored halos are able to grow in size significantly before much
        stellar stripping occurs. After a cuspy halo loses $50\%$ of its mass
        within the half-light radius due to tides, the satellite residing in the halo expands by $20\%$,
        whereas a satellite hosted by a similarly stripped cored halo expands by
        over a factor of $2$. Given that there is evidence that dwarf
        galaxies with stellar masses around $10^8~\msun{}$ live in cored dark
        matter halos \citep[e.g.][]{kuzio2008, oh2011a}, stellar expansion as a
        result of tidal heating appears to be a viable UDG formation scenario.

        Motivated by the environmental dependence of UDG formation as well as
        the predicted expansion of dwarf galaxies through tidal heating, we
        present a scenario in which UDGs are formed through tidal heating as
        dwarf galaxies orbit within a cluster. Using subhalos identified in the
        Illustris-dark simulation, we illustrate that the observed distribution of UDG
        properties are consistent with this scenario. In
        Section~\ref{sec:selection}, we describe the selection of cluster
        subhalos from Illustris-dark and the association of dwarf galaxies with
        subhalos at infall. In Section~\ref{sec:massloss} and \ref{sec:results},
        we outline our procedure for modeling the effect of tidal stripping and
        compare the properties of UDGs produced via tidal stripping with the
        observed cluster UDGs, before summarizing our conclusions in
        Section~\ref{sec:conclusions}.

	\section{Identification of Cluster Subhalos}
	\label{sec:selection}
        In order to explore the effect that tidal stripping has on the
        properties of dwarf galaxies, we utilize data from the Illustris-dark
        Simulation \citep{vogelsberger2014i,nelson2015}. Illustris-dark is a cosmological, dark
        matter-only simulation of $1820^3$ particles, each with a mass of
        $5.32\times10^{6}~h^{-1}~\msun{}$, in a $75~h^{-1}$~Mpc box, simulated
        using a Plummer-equivalent force softening length of $1.0$ comoving
        $h^{-1}~{\rm kpc}$. Illustris-dark was run with the following
        cosmological parameters: $h=0.704$, $\Omega_{\rm M}=0.2726$,
        $\Omega_{\rm \Lambda}=0.7274$, $n_s=0.963$, and $\sigma_8=0.809$. Halos with
        masses down to $10^8~\msun{}$ are identified using
        the Subfind algorithm \citep{springel2001}, and the merger
        trees are generated using the Sublink algorithm \citep{rodriguez-gomez2015}, 
        with halo properties output for $135$ time-steps logarithmically spaced in scale factor.
		 As we discuss in Section~\ref{sec:massloss}, the only subhalo properties we utilize directly 
		 from the simulations are the subhalo masses at infall and the subhalo orbits. These properties
		 are accurate to a high level ($<10$~kpc in position and $<10$~km~s$^{-1}$ in velocity;
		 \citealt{behroozi2013}).

        Given that halos in cluster environments are most likely to experience
        significant tidal stripping, we restrict our analysis to subhalos within
        $R_{200}$ of a cluster with $M_{200}>10^{14}~\msun{}$ at
        $z=0$. Here, and for the remainder of this work, $R_{\Delta}$ represents
        the radius within which the mean density of a halo is $\Delta$ times the
        critical density and $M_{\Delta}$ is the bound mass within
        $R_{\Delta}$. Subhalo virial masses $(M_{\rm vir})$ are defined according to the
        \cite{bryan1998} redshift-dependent criteria, which corresponds to
        $\Delta=97$ at $z=0$ for the Illustris-dark cosmology. Each subhalo is assigned
        a galaxy with a stellar mass based on the maximum value of $M_{\rm vir}$ before infall given by
        the stellar mass-halo mass relation for cluster environments from
        \cite{kravtsov2014}, with $0.2$ dex normally-distributed scatter
        \citep{behroozi2013a}.
		
		\subsection{Galaxy Sizes at Infall}
		\label{sec:infallsizes}
		
        In addition to the stellar mass determined from abundance matching,
        each galaxy is assigned a half-light radius at infall
        based on its stellar mass, following the observed size-mass relation
        from the GAMA survey \citep{driver2011,lange2015,liske2015}. Given that
        low-mass dwarf galaxies are quenched rapidly after falling into a
        massive host halo \citep[e.g.][]{fillingham2015,fillingham2016}, we use
        the size-mass relation for red galaxies as our initial condition:
        \begin{equation}
        \label{eqn:masssize}
          r_e=0.17\left(\frac{M_*}{\msun}\right)^{0.1}\left(1+\frac{M_*}{2.31\times10^{10}~\msun}\right)^{0.65}
          \; ,
        \end{equation}
        where $M_*$ is the stellar mass assigned to the subhalo at infall. The
        intrinsic scatter in $\log(r_e)$ at a given mass is taken to be $0.1$~dex, following
        measurements of the size-mass relation in CANDELS
        \citep{grogin2011,koekemoer2011,vanderwel2014}. While observations suggest significant size evolution
        of galaxy sizes with redshift among massive galaxies \citep{vanderwel2014}, the evolution of low-mass galaxies with $z$ is less clear.
        In contrast with observations of a significant decrease in galaxy sizes with $z$ among massive galaxies, hydrodynamic simulations find that
        dwarf galaxy sizes are primarily driven by cycles of star-formation and supernova feedback, and do not
        strongly evolve with $z$ \citep{chan2017,el-badry2016}.
        Furthermore, if we include the redshift-dependence of $r_e$ for massive galaxies from \cite{vanderwel2014} in our model,
        the $z=0$ dwarf galaxy population does not lie along the observed size-mass distribution of quiescent systems,
        so we do not include any redshift dependence in our starting size-mass relation.
        
        The applicability of this starting condition remains the largest source of uncertainty in our analysis.
        Understanding the processes involved in the transformation from star-forming dwarf-irregular galaxies
        (following the size-mass relation for blue systems), to quiescent dwarf-ellipticals (following
        the size-mass relation for red galaxies) is an active area of research 
        \citep[e.g.~][]{wheeler2017,kazantzidis2017,fattahi2018}.
        As quenching and tidal heating take place on similar timescales in this mass range ($\sim1$~Gyr), and the 
        vast majority of quiescent galaxies with $M_*=10^8~\msun{}$ live in dense 
        environments \citep{geha2012}, it is very difficult to disentangle the processes of tidal heating and 
        quenching that drive 
        the size evolution of these dwarfs.
        
        Regardless, there is evidence that most $M_*\sim10^8~\msun{}$ cluster dwarfs obey 
        Eqn.~\ref{eqn:masssize} before tidal stripping occurs.
        Firstly, most cluster dwarfs (including UDGs; \citealt{roediger2017,eigenthaler2018,yagi2016}) in this
        mass range are quenched, suggesting that any internal processes altering their sizes should have occurred 
        at some point before we observe them.
        Similarly, observations indicate that $\sim70\%$ of UDGs are nucleated \citep{yagi2016}, implying that they 
        were smaller
        systems with a more significant bulge before expanding. 
        Additionally, Eqn.~\ref{eqn:masssize}, in contrast with the size-mass relation for blue systems, produces a 
        substantial population of $M_*\sim10^8~\msun{}$ dwarf-ellipticals with
        small ($r_e<1$~kpc) sizes, in agreement with the observed size-mass relation for cluster dwarfs. In particular, the size distribution of $5\times 10^7-2\times 10^8\msun{}$ dwarfs produced through 
        this analysis following Eqn.~\ref{eqn:masssize} generally agrees with the size distribution of dwarfs of similar mass 
        identified in the Next Generation Fornax Survey \citep{munoz2015}.
        Lastly, if 
        the
        size-mass relation for blue galaxies is used as the initial condition for this analysis, the size distribution is flatter than 
        observed (see Sec.~\ref{sec:finalmstarrstar}).
        
        However, there is also evidence that some systems may follow the size-mass relation for blue galaxies 
        before they are stripped. If all dwarfs follow the quiescent size-mass relation before stripping, it is difficult 
        to fully reproduce either 
        the large velocity dispersions observed in some UDGs \citep[][see 
        Sec.~\ref{sec:dmhalos}]{beasley2016,vandokkum2016} or the abundance of extremely 
        extended ($>4.5$~kpc) UDGs (see Sec.~\ref{sec:finalmstarrstar}). If some UDG progenitors lie closer to the 
        size-mass relation for blue systems from the GAMA survey, such extreme UDGs are produced in our model.
        Furthermore, the quenching time associated with more massive dwarfs is significantly longer than the orbital time 
        of cluster subhalos \citep{wheeler2014}, suggesting that they continue to form stars as tidal stripping and 
        heating occur.
     
        The true relationship between the
		size and mass of dwarfs before they are stripped is likely in-between
		the star-forming and quiescent relations, particularly for systems with high $M_*$ at infall (that may take 
		longer to 
		quench) or systems that were accreted at high-$z$ (when galaxies may have been smaller in the field and clusters may have been less effective at quenching 
		infalling systems).
		Ultimately, we adopt the $z=0$ size-mass relation for quiescent systems
		as a straightforward, conservative estimate of the size distribution
		of cluster dwarfs before they are stripped, as it best matches the joint 
		size distributions of
		cluster UDGs and dwarf-ellipticals.

        \section{Modeling the Effect of Tides on Cluster Satellites}
        \label{sec:massloss}

	\begin{table}
          \begin{center}
            \begin{tabular}{|l|c|c|c|r|r|r|}
              \hline 
              $\gamma$ & $r_e/r_{\rm max}$ &  & $M_{\rm *}$ & $r_{e}$ & 
                                                                $V_{\rm max}$ & $r_{\rm max}$  \\ 
              \hline 
              $1$ (cusp) & $0.05$ &  $\alpha$ & $1.87$ & $0.47$ & $0.40$ & $-0.30$ \\ 
              $1$ & $0.05$ &  $\beta$ & $1.87$ & $0.41$  & $0.30$ & $0.40$\\ 
              $1$ & $0.05$ &  $\log_{10}(x_s)$ & $-2.64$ & $-2.64$  & $0$ & $0$\\ 
              $1$ & $0.1$ &  $\alpha$ & $1.80$ & $0.50$ & $0.40$ & $-0.30$ \\ 
              $1$ & $0.1$ &  $\beta$ & $1.80$ & $0.42$ & $0.30$ & $0.40$\\ 
              $1$ & $0.1$ &  $\log_{10}(x_s)$ & $-2.08$ & $-2.08$ & $0$ & $0$\\
              \hline 
              $0$ (core) & $0.05$ &  $\alpha$ & $2.83$ & $-0.25$ & $0.40$ & $-1.30$\\ 
              $0$ & $0.05$ &  $\beta$ & $2.83$ & $-0.25$ & $0.37$ & $0.05$ \\ 
              $0$ & $0.05$ &  $\log_{10}(x_s)$ & $-3.12$ & $0$ & $0$ & $0$\\
              $0$ & $0.1$ &  $\alpha$ & $0.330$ & $2.05$ & $0.40$ & $-1.30$\\ 
              $0$ & $0.1$ &  $\beta$ & $0.140$ & $2.05$ & $0.37$ & $0.05$\\
              $0$ & $0.1$ &  $\log_{10}(x_s)$ & $0$ & $0$ & $0$ & $0$\\ 
		 
              \hline 
            \end{tabular}
          \end{center}
          \caption{The $\alpha$, $\beta$, and $x_s$ parameters, for a given
            $\gamma$, $r_{e}/r_{\rm max}$, utilized in Equation~\ref{eqn:trackeqn}
            to model the changes in stellar mass, half-light radius, halo 
            $V_{\rm max}$, and halo $r_{\rm max}$ as a consequence of
            tidal stripping. Note that the $r_e$ and $M_*$ tracks, generated among subhalos in the Aquarius-A2 
            simulations, take the ratio of the final mass within
            $r_{\rm max}$ to the initial mass within $r_{\rm max}$ as $x$ in Equation~\ref{eqn:trackeqn}, whereas 
            the $r_{\rm max}$
            and $V_{\rm max}$ tracks, generated by simulating multiple subhalo orbits within a Galactic potential, 
            take ratio of the initial mass within the tidal radius to the final mass within the tidal radius as $x$ (see 
            Sec.~\ref{sec:massloss}).
            } 
          \label{tab:abtable}
        \end{table}

		        \begin{figure*}
	\centering
	\includegraphics[width=0.8\linewidth]{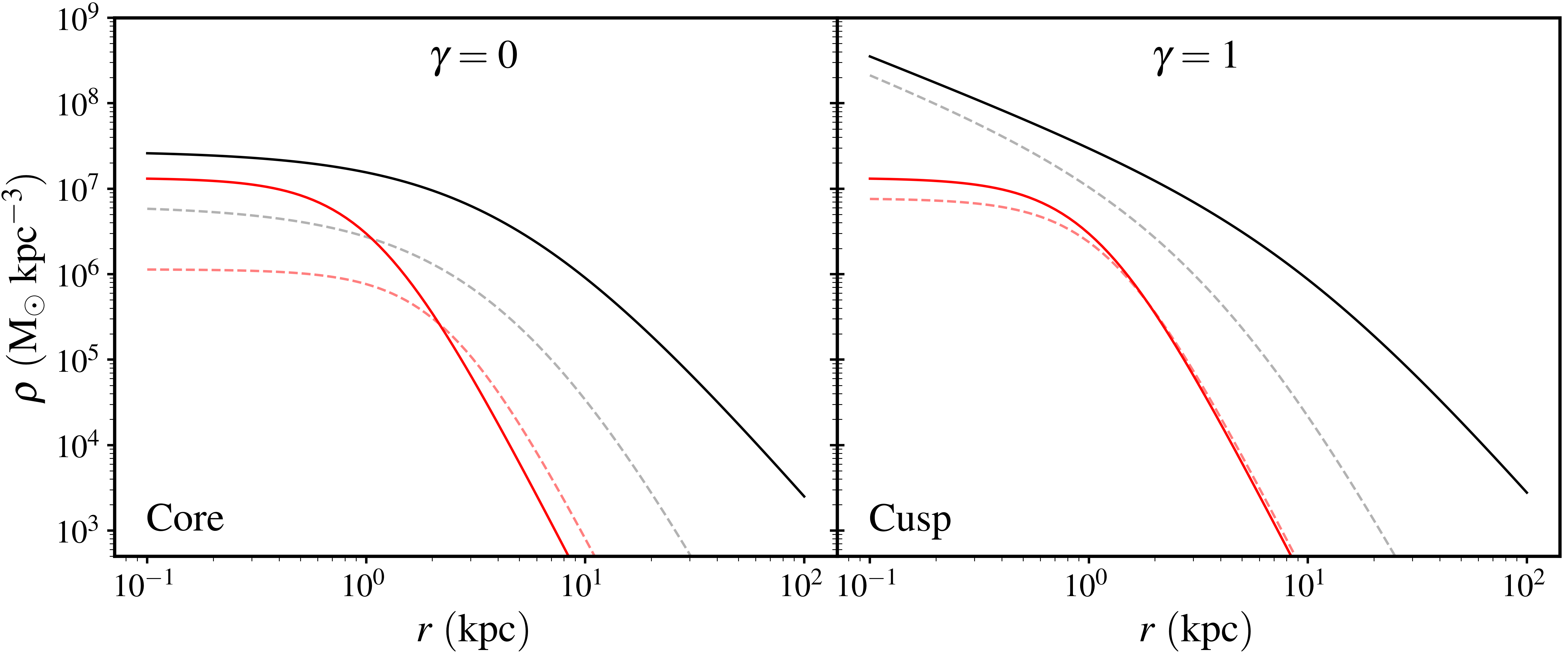}
	\caption{The effect of tidal stripping on galaxies in both cuspy
		(\emph{right}) and cored (\emph{left}) dark matter halos. The black
		and the red lines represent the dark matter and stellar
		density distributions for a representative system at infall. After the
		galaxy is tidally stripped, the resulting stellar and dark matter
		density distributions are shown as dashed light red and grey lines,
		respectively. At infall, the galaxy has a dark matter mass of
		$3.6\scinot{10}$~\msun{} and stellar mass of
		$7.5\scinot{7}$~\msun{}, with a stellar half-light radius of
		$1.1$~kpc. After three pericentric passage within $350$~kpc of a
		cluster with $M_{200}=2.4\times10^{14}~\msun{}$, the halo
		loses $98\%$ of its mass. The cored and cuspy systems have similar stellar masses
		at $z=0$ ($6.7\scinot{7}$~\msun{} and $6.9\scinot{7}$~\msun{},
		respectively); however, the half-light radius of the cored system
		grows to $2.4$~kpc, while the half-light radius of the cuspy system
		only grows slightly to $1.3$~kpc.}
	\label{fig:masslosscluster}
\end{figure*}
        In order to model the effect that tides have on the size and
        stellar mass of dwarf galaxies, we make use of results from
        \cite{errani2015}. They find that the change in stellar mass
        and stellar half-light radius of a tidally-stripped and heated galaxy can be
        parameterized as a function of the total amount of mass lost within the
        half-light radius and the inner slope of the dark matter density
        profile. To improve the applicability of this tidal parameterization to
        cosmological simulations, updated tidal tracks have been created using
        satellites drawn from the Aquarius-A2 merger tree \citep{springel2008} in which the stellar
        population is treated as a collection of collisionless tracer particles \citep{errani2017,errani2018} and the dark-matter halo is set up initially as a \cite{dehnen1993} profile. 
        These updated tracks \citep[originally introduced in][]{penarrubia2008} are
        parameterized as a function of the mass loss experienced within the
        radius corresponding to the maximum circular velocity ($r_{\rm
          max}$).
      Both the change in stellar mass and stellar half-light radius
        are expressed using the functional form:
        \begin{equation}
          \label{eqn:trackeqn}
          g(x)=\frac{\left(1+x_s\right)^{\alpha} x^{\beta}}{\left(x+x_s\right)^{\alpha}} \; ,
        \end{equation}
        where $g(x)$ represents either $M_{\rm *,~final}/M_{\rm *,~initial}$ or
        $r_{e,~{\rm final}}/r_{e,~{\rm initial}}$ and $x$ is
        $M(r<r_{\rm max})_{z=0}/M(r<r_{\rm max})_{\rm infall}$, with
        $r_{\rm max}$ evaluated at $z=0$ and infall, respectively.
        As given in Table~\ref{tab:abtable}, the values for $\alpha$, $\beta$, and $x_s$ are determined by fitting $g(x)$ to
        systems with $<99\%$ mass loss and depend on the inner slope of the dark
        matter density profile, $\gamma$, as well as the ratio between $r_e$ and $r_{\rm max}$. 
        Although a second-order
        dependence of the tracks on the shape of the stellar distribution is apparent, a Plummer density
                distribution \citep{plummer1911} for the stellar profile of all satellites is assumed
        given the limited information regarding the stellar
        distributions of UDGs. Given that the $r_e/r_{\rm max}$ values within our
        sample span a wide range (the $5-95$ percentile range is $0.04$ to $0.28$), we logarithmically interpolate 
        (and extrapolate
        when required) between the
        distinct mass-loss tracks to determine $g(x)$ for each halo, with the condition that $M_{*,~{\rm 
        final}}/M_{*,~{\rm initial}}\le1$. Although this requires significant extrapolation, the tracks only weakly 
        depend on $r_e/r_{\rm max}$ (the median absolute deviation of the interpolated/extrapolated tracks from 
        the measured tracks due to this correction is $1\%$ both in terms of $M_{\rm *,~final}/M_{\rm *,~initial}$ or
        $r_{e,~{\rm final}}/r_{e,~{\rm initial}}$); the adoption of a constant $r_e/r_{\rm max}$ does not significantly 
        impact our results.

		\subsection{Mass Loss Experienced by Subhalos}
		This analysis relies on an accurate measurement of the amount of mass
		lost as a result of tides for each subhalo. As Illustris-dark was run with cold collisionless dark matter particles, the simulation only produces cuspy halo density profiles. Thus, the mass loss experienced by halos in these simulations
		is insufficient to model the tidal effects experienced by cored dwarfs.
		Moreover, recent work by \citet{vandenbosch2016},\citet{vandenbosch2017}, and \citet{vandenbosch2018} show that up to $80\%$
		of tidal stripping observed in the Bolshoi simulation \citep{klypin2011} could be the result of numerical
		artifacts rather than physical stripping. In lieu of the mass loss as
		measured in Illustris-dark, we use the procedure outlined in
		\citet{penarrubia2010} to determine the change in mass and shape of dark
		matter halos as a result of tidal stripping. This procedure, briefly
		outlined below, only relies on knowledge of the subhalo orbit and the
		initial halo profile, which are
		subject to significantly less uncertainty than the measured subhalo
		properties at $z=0$, to determine the amount of stripping.
		
		For each subhalo, we track its orbit from infall to the first
			pericentric passage. At pericenter, the amount of bound mass is calculated
			as the total mass within the tidal radius. Following \cite{vandenbosch2018},
			we calculate the tidal radius as
			\begin{equation}
			\label{eqn:rteqn}
			R_{t,~1}=\left(\frac{G~ m_{\rm
					sat}(R_{t,~1})}{\omega^2-\frac{d^2\Phi}{dr^2}}\right)^{1/3} \; ,
			\end{equation}
			or
			\begin{equation}
			\label{eqn:rteqn}
			R_{t,~2}=D\left(\frac{m_{\rm sat}(R_{t,~2})}{M_{\rm host}(D)}\right) \; ,
			\end{equation}
			where $m_{\rm sat}(R_t)$ is the mass of the galaxy within the tidal radius (including both the dark-matter 
			and stellar component), $D$ is the distance between the satellite and the host cluster, $M_{\rm host}(D)$ is the mass of the host cluster within $D$, and
			$\omega$ is the angular velocity of the subhalo, measured directly from
			the simulation. The second derivative of the host potential at pericenter, $\frac{d^2\Phi}{dr^2}$,
			is calculated assuming an NFW profile, using the
			mass and scale radius of the cluster identified in the halo
			catalogs.
			Following \cite{vandenbosch2018}, we chose $R_{t,~1}$ or $R_{t,~2}$ as the tidal radius depends on the angular momentum of the orbit normalized by the angular momentum of a circular orbit with the same energy $(\eta)$. If $\eta$ is greater than $0.75$, the tidal radius is taken to be $R_{t,~1}$; otherwise we use $R_{t,~2}$.

		The mass within the tidal radius at pericenter divided by the maximum
		virial mass of the subhalo before infall $(M_{\rm peak})$ is then used as input for the
		tidal tracks (Eqn.~\ref{eqn:trackeqn}, using the $\alpha$, $\beta$, and $x_s$ values
		from Table~\ref{tab:abtable}) to determine the change in $V_{\rm max}$
		and $r_{\rm max}$ of the halo. For halos that do not have a pericenter before $z=0$, 
		the mass within the tidal radius at $z=0$ divided by $M_{\rm peak}$ is used to determine
		the changes in $r_{\rm max}$ and $V_{\rm max}$. If a halo experiences $>10\%$ mass loss,
		the log-slope of the outer density profile is changed from $3$ to $5$ to
		match the observed outer density profiles of tidally-stripped galaxies
		\citep{penarrubia2008,penarrubia2009}. This procedure is repeated for
		each pericentric passage to establish the $z=0$ profile. This analysis has been shown to
		reproduce the amount of mass loss in halos experiencing tidal stripping
		\citep[see Appendix A of][]{penarrubia2010}. Among halos with more than $10000$ particles at infall in the Illustris-dark simulation,
			the mass loss produced through this technique reproduces the simulated mass loss well (with $\left<{\rm Log}[M_{z=0,~{\rm model}}/M_{z=0,~{\rm sim}}]\right>=0.28$ and $\sigma_{{\rm Log}[M_{z=0,~{\rm model}}/M_{z=0,~{\rm sim}}]}=0.43$).

		In line with results from cosmological simulations \citep{klypin2011},
			cuspy subhalos are modeled with NFW profiles \citep{navarro1997}:
			\begin{equation}
			\label{eqn:nfw}
			\rho(r)=\frac{\rho_s}{\left(\frac{r}{r_s}\right) \left(1+\frac{r}{r_s}\right)^{2}}
			\; ,
			\end{equation}
			where $\rho_s$ is the normalization and $r_s$ is the scale radius. For a given halo,
			$r_s$ and $\rho_s$ are assigned based on the $M_{200}$ at infall and the 
			redshift-dependent mass-concentration relation from \cite{prada2012} with $0.16$~dex scatter \citep{diemer2015}.
		Cored halos are modeled with cored NFW profiles as:
		\begin{equation}
		\label{eqn:corenfw}
		\rho(r)=\frac{\rho_s}{ \left(1+\left(\frac{r}{r_s}\right)\right)^{3}} 
		\; .
		\end{equation}
		For the cored halos, $\rho_s$ and $r_s$
		are taken so that the $V_{\rm max}$ and $r_{\rm max}$ of the cored halo match their corresponding
		values for the corresponding cuspy halo. 
		Although the physical processes behind core generation are under extensive debate, this model is
		generally consistent with observations, as well as SIDM models with low 
		($\sim 1~{\rm cm^2/g}$) cross sections. 
		Other models, motivated by the fact that supernova feedback can lower the dark matter density in the centers of halos \citep{dicintio2014,madau2014,pontzen2014,read2016,fitts2017,zolotov2012}, suggest smaller cores, closer to the galaxy
		half-light radius.
		We test the impact of different profile shapes on our analysis in Appendix~\ref{sec:appendix}, including the \cite{dicintio2014} profile; however as the mass profiles of halos with feedback-generated cores are not too different from our canonical halo, particularly for $r>r_s$ where most tidal radii occur, our conclusions are not very sensitive to the exact shape of cored halos at infall. Additionally, we note that the although \cite{errani2018} tracks were modeled with a \cite{dehnen1993} profile (with an outer slope of $4$ instead of $3$), they should be applicable for NFW profiles given that the stellar component is well within the inner region of our halos for our systems.
		Thus, we remain agnostic to the mechanism producing cored halos, so long as the core is formed before the halo's infall onto the cluster.
		
		\begin{figure*}
			\centering
			\begin{tabular}{cc}
				\includegraphics[width=.5\linewidth]{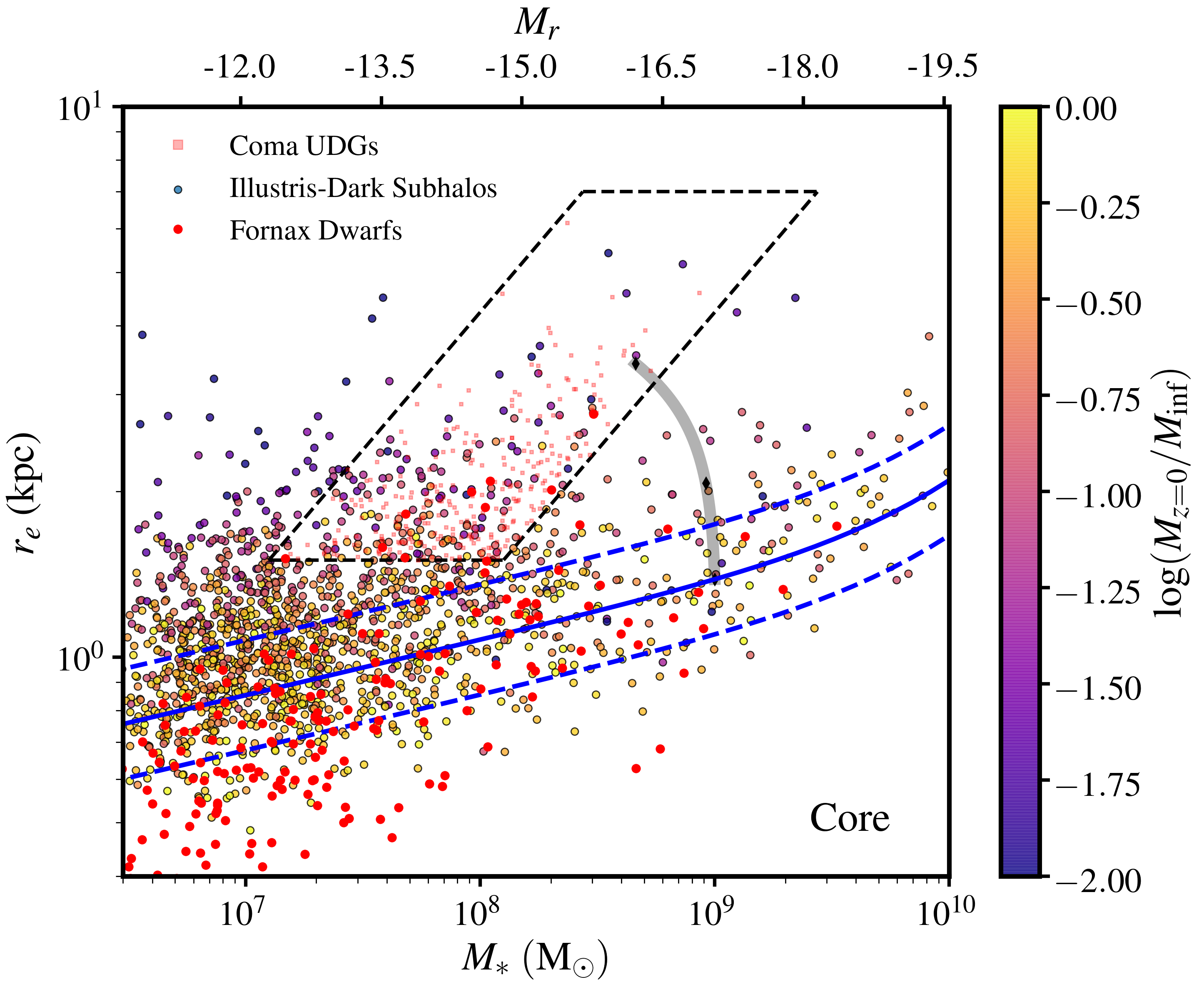}\label{fig:coremagsize}
				& 
				\includegraphics[width=.5\linewidth]{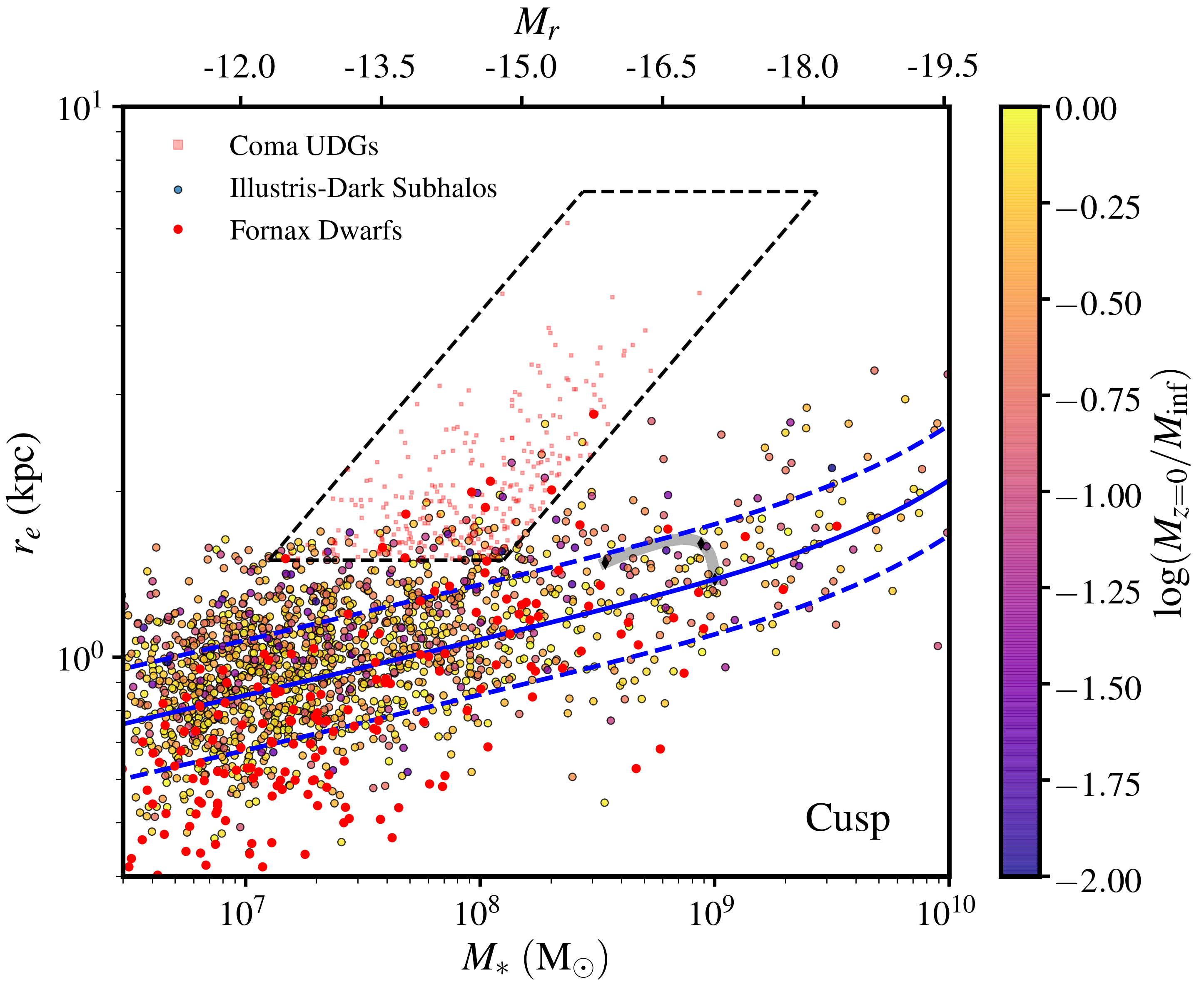}\label{fig:cuspmagsize}
			\end{tabular}
			\caption{The relationship between stellar mass and half-light radius
				for both observed systems in the Coma cluster (red squares) and
				simulated objects (colored points) in cored (\emph{left}) and cuspy
				(\emph{right}) halos within a $M_{200}=2.4\scinot{14}~\msun$ cluster. 
				The degree of mass loss experienced between
				infall and $z=0$ for cored and cuspy systems is illustrated by the
				coloring of each point. Blue solid and dashed lines indicate the
				starting mass-size relation and the associated $1\sigma$ scatter
				that simulated systems are drawn from at infall, while the black
				dashed lines illustrate the UDG selection criteria. The thick grey
				line shows the evolution of a $10^9$~\msun{} galaxy experiencing up
				to $99$~percent mass loss, with black diamonds denoting the location
				of the galaxy at $0$, $90$, and $99$~percent mass loss (evolving
				from low to high $r_e$, respectively).
				Also shown as red points are satellites in the Fornax cluster \citep{munoz2015,eigenthaler2018}.
				Overall, cuspy halos do not
				experience the required mass loss to generate a substantial number
				of UDGs, whereas systems in cored subhalos produce UDGs with a range
				of sizes and stellar masses that broadly agree with observed UDG
				samples.}
			\label{fig:magsize}
		\end{figure*}

		Models including SIDM as well as baryonic 
		effects fit Eqn.~\ref{eqn:corenfw} well but evolve over time in the field, such that halos that fell into the cluster at earlier times have smaller cores \citep{kamada2017}.
		While our procedure may overestimate the core sizes for the $54\%$ of subhalos that fell in before $z=1$ (at which point core sizes are within $\sim30\%$ of their $z=0$ sizes; \citealt{kaplinghat2016}),
			analysis in Appendix~\ref{sec:appendix} demonstrates that our results are not particularly sensitive to the core size at infall, as long as the galaxy evolves along the cored tracks.
		
     While this procedure is appropriate for the majority of low-mass subhalos (below $\sim 1\scinot{11}$~\msun),
     it may be inappropriate for higher mass subhalos. The simulations producing the \cite{errani2018} tracks 
     are exclusively set up in systems where the dark-matter density is higher than the stellar density 
     at all radii. This condition does not apply to all systems in our model, however, with many higher-mass systems exhibiting baryon-dominated centers. Although this is unlikely to affect the 
     overall evolution of the dark-matter halo in response to tides (which is sensitive to the mass profile far 
     away from the stellar component) it may affect the response of the stellar profile to tidal heating (which is 
     sensitive to the potential felt by the stellar component). To model this possible effect, we treat systems with 
     baryon-dominated centers as intermediate cases between cores and cusps, with $\gamma$ between $0$ 
     and $1$. A conservative estimate for the appropriate value of $\gamma$ is determined using the ratio of stellar mass to dark-matter mass within the stellar half-light radius. If this quantity $(\delta_{1/2})$ is less than $1.0$, $\gamma$ is set to $0$. If $\delta_{1/2}$ 
     is greater than $2.0$, $\gamma$ is set to $1$. For values in between $1.0$ and $2.0$, $\gamma$ is set to 
     $\delta_{1/2}-1$.
     The $M_{\rm *,~final}/M_{\rm *,~initial}$ and $r_{e,~{\rm final}}/r_{e,~{\rm initial}}$ tracks are interpolated to the 
     appropriate value of $\gamma$ and the stellar profile is evolved according to the tracks for this 
     intermediate $\gamma$. Given that the shape of the gravitational potential in the central regions of baryon-dominated 
     systems in cored halos more resembles that of a cored halo than a cuspy one (as UDGs, and dwarf galaxies in general, have shallow stellar density profiles \citealt{faber1983,vandokkum2015,yagi2016}), this procedure provides a 
     method for conservatively estimating the change in the stellar profile of baryon-dominated systems.

     The combination of $M_{\rm *,~final}/M_{\rm *,~initial}$ and
     $r_{e,~{\rm final}}/r_{e,~{\rm initial}}$ determined from
     Equation~\ref{eqn:trackeqn} and the initial $M_*$ and $r_e$ values
     assigned in Section~\ref{sec:selection} establish the final $M_*$ and
     $r_e$ values that are observed at $z=0$. An example of the transition
     between the infall profile and the $z=0$ profile is illustrated in
     Figure~\ref{fig:masslosscluster}.
     As the inner region of the cuspy halo is relatively unaffected by tidal stripping, the stellar profile is not 
     significantly affected. On the other hand, stripping and heating are able to penetrate to the center of the 
     cored halo, allowing for tidal heating in the central regions and an expansion of the stellar population.

		In summary, our analysis proceeds as follows:
		\begin{enumerate}[leftmargin=0.3cm]
			\item Identify all subhalos around massive clusters in Illustris-dark at $z=0$.
			\item Generate model cored and cuspy halo density profiles for each subhalo, based on $M_{\rm peak}$ and the cosmological mass-concentration relation.
			\item Assign each halo a stellar mass (based on abundance-matching) and half-light radius (based on the dwarf-elliptical size-mass relation) before its infall onto the cluster.
			\item Determine the degree of dark-matter stripping using the subhalo orbit and redshift-dependent host properties taken directly from the simulation. Concurrently change the subhalo density profiles using the \cite{penarrubia2010} tracks.
			\item Based on the amount of mass loss experienced by $z=0$, use the \cite{errani2018} tracks to derive the final stellar mass and half-light radius.
		\end{enumerate}

	\section{Properties of Tidally-Stripped UDGs}
	\label{sec:results} 
	\subsection{Stellar Masses and Half-Light Radii}
	\label{sec:finalmstarrstar}
	
	Given the procedure outlined in Section~\ref{sec:massloss}, we determine
        the stellar masses and half-light radii of cluster satellites in the
        Illustris-dark simulation, following tracks for both cored and cuspy
        dark matter halos. Figure~\ref{fig:magsize} illustrates the resulting
        satellite population of a $1.2\times10^{14}~\msun{}$ cluster alongside
        the UDGs observed in the Coma cluster ($M_{200}=2\times10^{15}~\msun{}$)
        from \cite{yagi2016} and dwarf-ellipticals from the Next Generation Fornax Survey\footnote{The virial
        mass of Fornax is $M_{200}=7\scinot{13}~\msun{}$ \citep{schuberth2010}.} \citep{munoz2015,eigenthaler2018}.
        Galaxies in cored dark matter halos, which
        experience significantly more mass loss than those in cuspy halos, span
        a broad range of sizes, whereas galaxies in cuspy halos largely fall
        along the field size-mass relation.

	In order to compare the properties of our simulated, tidally-stripped
        UDGs with observational samples, we select UDGs as galaxies with
        $\Sigma_*=M_*/(\pi r_e^2)$ between $1.73\times 10^6$ and
        $17.3\times 10^6$~\msun{}~kpc$^{-2}$ and $r_e$ between $1.5$ and $7$~kpc
        at $z=0$. The $r_e$ criterion is in line with existing UDG definitions
        \citep{koda2015,vanderburg2016}, and the surface brightness criterion
        corresponds to $24<\left< \mu \right>_e <26.5$,\footnote{Cosmological
          surface brightness dimming at the distance of Coma, $\sim~0.1~\msa$,
          is taken into account when converting observed $\left<\mu \right> _e$
          to $\Sigma_*$.} where $\left< \mu \right>_e $ is the average surface
        brightness within $r_e$ assuming an $r$-band mass-to-light ratio of
        $1.96$. This mass-to-light ratio, taken from \cite{zibetti2009} assuming
        $g-r=0.689$,
        is subject to significant uncertainty due to a lack of deep near-IR
        photometry.\footnote{We convert the CFHT colors to the SDSS filter set
          according to the following transformations: \\
          \indent $g_{\rm CFHT} = g_{\rm SDSS} - 0.153 (g_{\rm SDSS} - r_{\rm
            SDSS})$ \\
          and \\
          \indent $r_{\rm CFHT} = r_{\rm SDSS} - 0.024 (g_{\rm SDSS} - r_{\rm SDSS})$.}
        However, cluster UDGs appear to have optical colors consistent with the
        red sequence \citep{yagi2016}, so this assumption is not likely to significantly affect
        our results.

\begin{figure}
        	\centering
        	\includegraphics[width=1\linewidth]{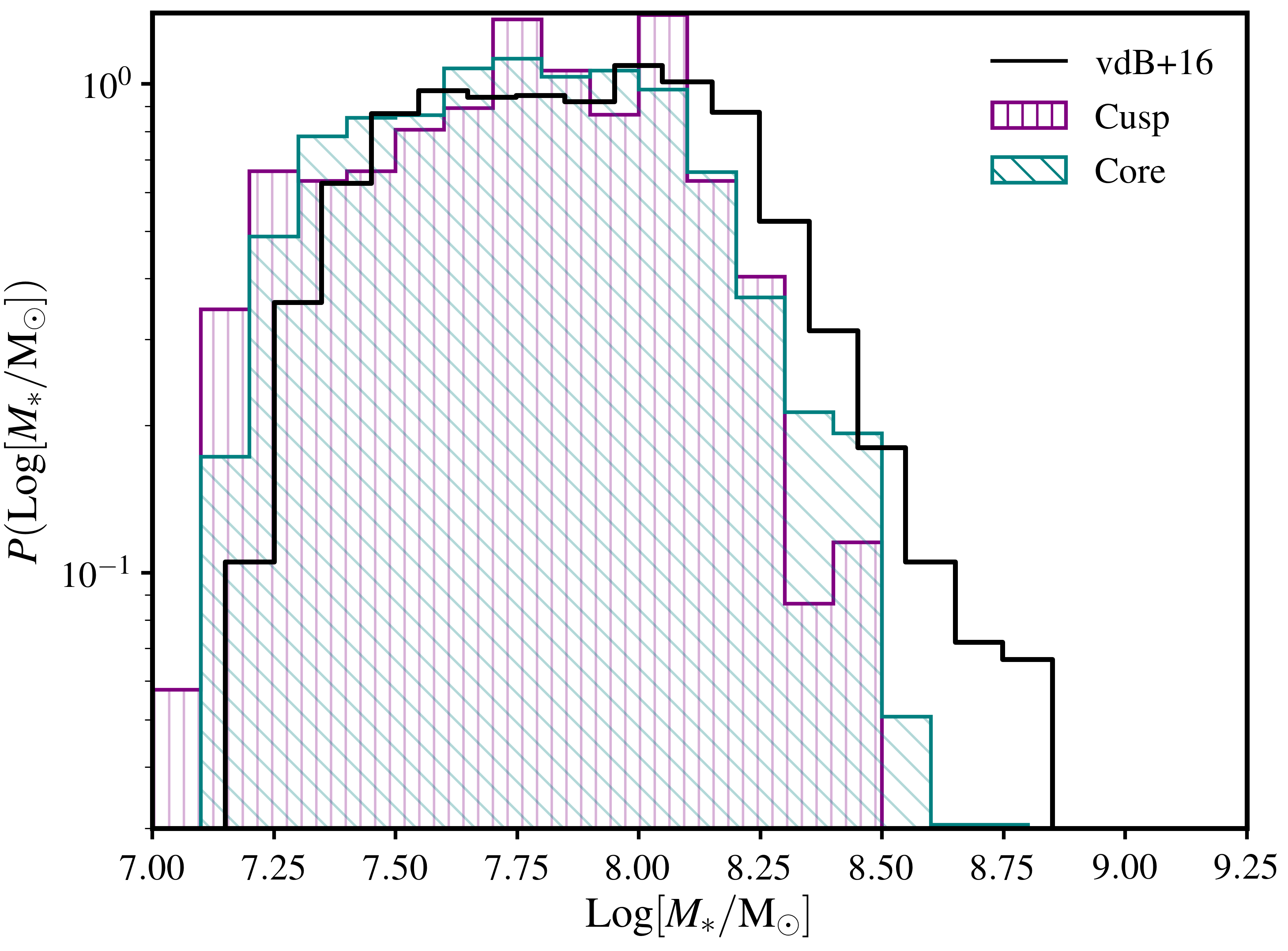}
        	\caption{The relative mass distribution of tidally-stripped UDGs
        		within cored (teal) and cuspy (purple) halos compared with the
        		observed UDG mass function from \protect \cite{vanderburg2016}
        		(black). Both cuspy and cored distributions match the observed
        		distribution remarkably well, peaking between $10^{7.5}$ and
        		$10^{8}$~\msun{}. As shown in Figure~\ref{fig:magsize}, the sizes of
        		galaxies within this mass range are only a few tenths of a dex away
        		from the UDG criteria, such that little-to-no transformation is
        		necessary to transform them into UDGs. Additionally, the large sizes
        		of UDGs in cored profiles allow for a population of high-mass UDGs,
        		similar to observations.}
        	\label{fig:masshist}
        \end{figure}

	The stellar mass distribution of tidally-stripped UDGs, illustrated in
        Figure~\ref{fig:masshist}, peaks at $10^8~\msun{}$ for both cuspy and
        cored halos, consistent with observations \citep{vanderburg2016}. As
        shown in Figure~\ref{fig:magsize}, the UDG criteria at this mass is only
        $60\%$ ($2.3\sigma$) away from the field size-mass relation, such that
        $\sim1\%$ of $10^8~\msun{}$ galaxies start off as UDGs simply due to
        scatter in the size-mass relation. Nevertheless, UDGs in cored halos
        distinguish themselves from those produced in cuspy halos, as they
        include a significant population of high-mass systems, in agreement with
        observations, whereas UDGs in cuspy halos do not. Because of the criteria
        used to define UDGs, these high-mass UDGs can only be created through
        significant tidal heating of more massive satellites (see Fig.~\ref{fig:coremagsize})
        in cored halos.
        
\begin{figure}
               	\centering
               	\includegraphics[width=1\linewidth]{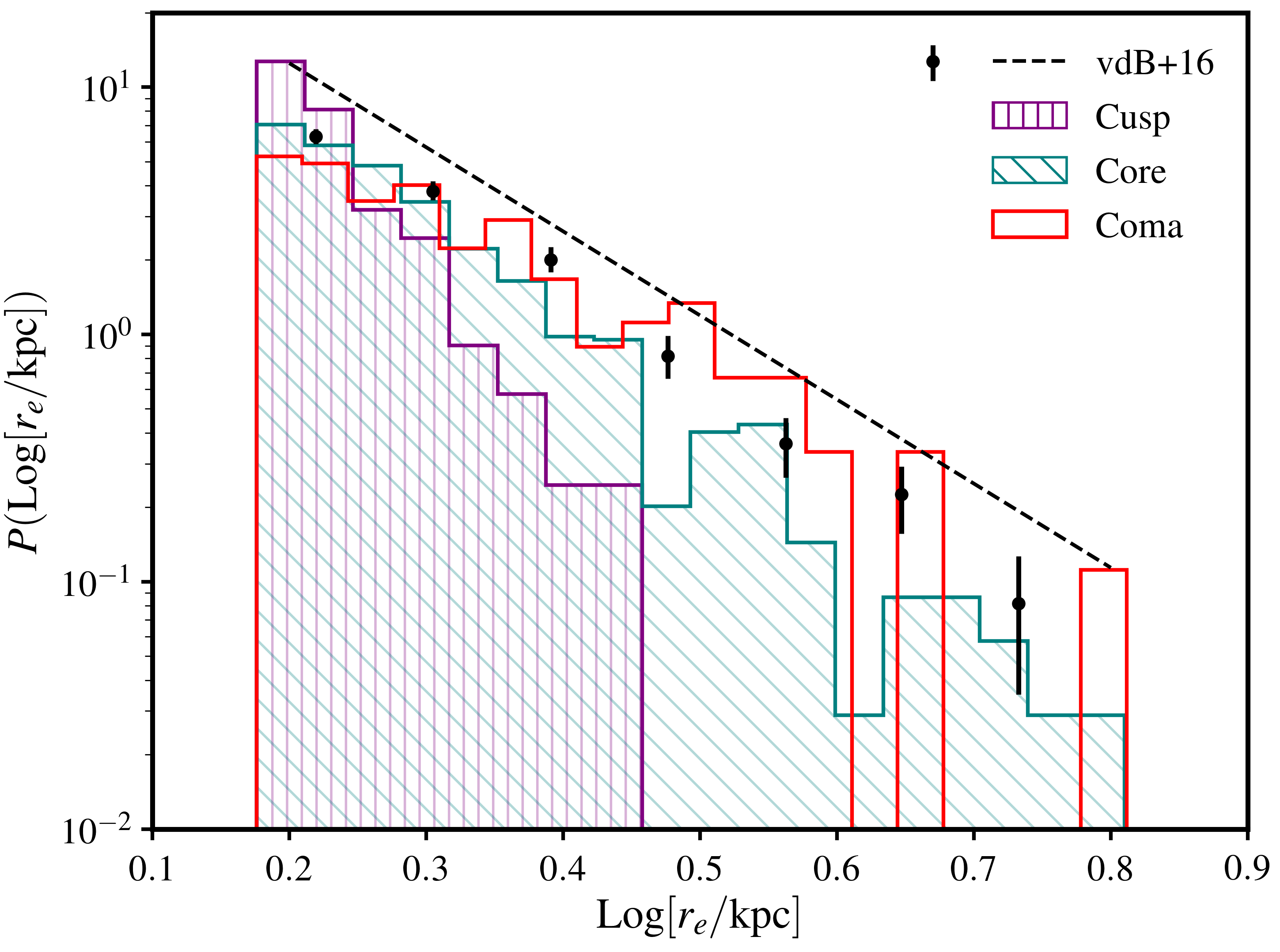}
               	\caption{The distribution of $r_e$ for tidally-stripped UDGs within both
               		cuspy (purple) and cored (teal) subhalos. Also shown, as black points, 
               		is the size distribution
               		of UDGs within nearby clusters (and the fit to that distribution as the black line;  \protect \cite{vanderburg2016}),
               		as well as the size distribution of Coma UDGs from
               		\protect \cite{yagi2016} as the red histogram. The high-$r_e$ tail of the observed
               		distribution is reproduced for UDGs in cored subhalos, but not for
               		UDGs in cuspy subhalos. In particular, tidal stripping of galaxies in
               		cuspy dark matter halos is not able to produce the population of large
               		($r_e\gtrsim3$~kpc) UDGs. Although our model appears to underpredict the
               		abundance of extremely extended ($>4.5~{\rm kpc}$) systems, such systems
               		would be expected in our model if they followed the size-mass relation
               		for star-forming systems before stripping (see Sec.~\ref{sec:infallsizes}).}
               	\label{fig:rehist}
               \end{figure}

	On the other hand, the distributions of the half-light radii of UDGs
        produced in cored and cuspy halos differ more
        substantially. Figure~\ref{fig:rehist} shows that the observed
        high-$r_e$ tail of the size distribution is reproduced in cored, but not
        in cuspy UDGs.
        While the size distribution of UDGs in cored halos
        follows $n\propto r^{-4.0}$, which is slightly steeper than the
        $n\propto r_e^{-3.4\pm0.19}$ found by \cite{vanderburg2016}, the size
        distribution of UDGs in cuspy halos falls much more steeply than the
        observations (following $n \propto r_e^{-7.6}$). The fraction of cored UDGs with $r_e$ above $3$~kpc is $4.7\%$, and the fraction 
        of UDGs with $r_e$ above $5$~kpc is $0.4\%$, compared with the observed fractions of $16.8\%$ and 
        $3.1\%$ respectively. 
        While our model does not quite produce the same population of extremely extended ($>4.5~{\rm kpc}$)
        systems found in observations, this population can easily be explained if we assume that they follow the 
        size-mass relation
        for blue systems before stripping occurs. 
              
        However, under the unreasonable assumption that \emph{all} systems follow the star-forming 
        size-mass relation before stripping (see Sec.~\ref{sec:infallsizes}), the size distribution is significantly flatter 
        than observed for both cored ($n\propto r^{-2.3}$) \emph{and} cuspy ($n\propto r^{-2.8}$) halos and does not
        reproduce the observed dwarf-elliptical size distribution.
        Overall, the ability of
        tidal stripping of satellites in cored halos to reproduce the UDG size 
        distribution from $1.5$ to $>3$~kpc is strong evidence in favor of this
        formation mechanism.

	\subsection{Stellar Populations}
        Given that tidally-stripped UDGs must have orbited the host cluster long
        enough to experience multiple pericentric passages, our analysis
        suggests that UDGs are primarily composed of old stellar populations.
        In particular, the largest UDGs, which in our formation scenario are
        formed after significant tidal stripping, should host significantly older
        stellar populations than smaller UDGs and non-UDG dwarf-ellipticals.
        As seen in Figure~\ref{fig:magsize}, the largest
        ($r_e \gtrsim 3$~kpc) UDGs have lost over $90$~percent of their
        dark matter mass. On the other hand, smaller UDGs can be formed without
        significant mass loss, and may include younger stellar populations. If
        satellites are taken to be quenched $\sim1$~Gyr after infall
        \citep{fillingham2015}, our analysis predicts that the average stellar
        age of UDGs with $r_e$ between $1.5$ and $3$~kpc is $4.8$~Gyr with the $10-90$ percentile spread $0.3-8.0$~Gyr, but the age of UDGs with $r_e$ between
        $4.5$ and $6$~kpc is $7.8$~Gyr, with the $10-90$ percentile spread $6.2-9.1$~Gyr. This prediction is consistent with the $\sim10$~Gyr ages
        measured in large UDGs in Coma \citep{gu2017,pandya2017,ruiz-lara2018,ferre-mateu2018}, however, a more complete exploration of the size-age parameter space will provide better constraints on this model.

                 \begin{figure*}
               	\centering
               	\includegraphics[width=1\linewidth]{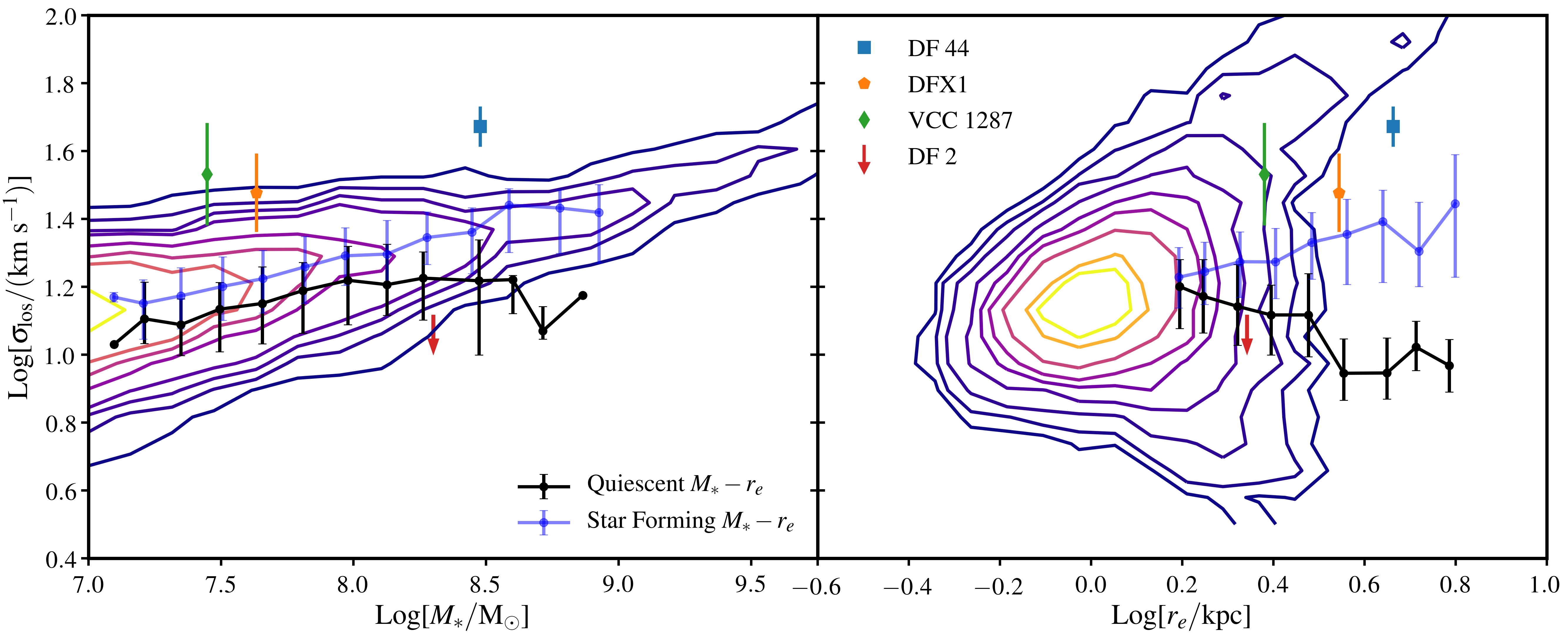}
               	\caption{The relationship between line-of-sight velocity dispersion and final stellar mass (\emph{left}) 
               		and final half-light radius (\emph{right}), highlighting the field population as contours and the 
               		UDG population 
               		as 
               		points, with the $25-75\%$ range illustrated with error bars. Black points illustrate the UDG 
               		population generated by our model
               		assuming the size-mass relation for quiescent systems before stripping, whereas light blue 
               		points are generated assuming the star-forming size-mass relation before 
               		stripping.
               		Although UDGs are biased toward slightly lower velocity dispersions as a consequence of tidal stripping, they 
               		otherwise mimic the field relationships.
               		Similarly, as UDGs which started on the star-forming size-mass relation have experienced less stripping than those starting on the quiescent size-mass relation, their dispersions are somewhat higher for high-mass UDGs.
               		Also shown is the velocity dispersions of DF~44 (shown as the blue square), the globular cluster velocity dispersions VCC~1287 and DFX1
               		(shown as the 
               		orange pentagon and green diamond), and the upper limit on the globular cluster velocity dispersion of DF2 (as the red arrow).
               		The existence of very dark-matter dominated objects like DF~44 and 
               		VCC~1287 may be explained if they lie slightly above the size-mass relation for star-forming systems (see 
               		Sec.~\ref{sec:dmhalos}).}
               	\label{fig:dispersions}
               \end{figure*}

        Recognizing that our simulated UDGs are born in low-mass halos
        ($\sim10^{10}-10^{10.5}~\msun{}$), we expect them to have low
        metallicities consistent with the classical low-mass dwarfs
        ($[{\rm Fe}/{\rm H}]\sim-1.1$, \citealt{kirby2013}). This aligns with
        spectral \citep{kadowaki2017, gu2017,ruiz-lara2018,ferre-mateu2018} and color-based
        \citep{roman2017iso,pandya2017} measurements of the stellar metallicities of UDGs.

	\subsection{Dark Matter Halos}
	\label{sec:dmhalos}

        Of particular interest when considering the formation of UDGs are the
        dark matter halos that host these galaxies. For example, the high
        stellar velocity dispersion observed for DF~44 --- implying a central
        dark matter fraction of $>98\%$ --- motivated the idea that UDGs are
        primarily formed in Milky~Way-like halos that failed to form a
        substantial stellar mass \citep{vandokkum2015b}. In our proposed
        scenario, however, UDGs are formed in similar dark matter halos as
        typical dwarf ellipticals, characterized by stellar velocity dispersions
        of $\sim15$~km~s$^{-1}$ within the half-light radius today. Although
        tidally-stripped UDGs experience significant dark matter mass loss, the
        mass loss is primarily in the halo outskirts (see
        Fig.~\ref{fig:masslosscluster}), creating UDGs which contain similar amounts of dark matter within the 
        half-light radius as typical dwarf galaxies.
        The average line-of-sight velocity
        dispersion of cored UDGs is $14$~km~s$^{-1}$, with the $10-90$ percentile range extending from $9$
        to $23$~km~s$^{-1}$. 
        
        For our analysis, we compute dispersions from the line-of-sight virial theorem \citep{merrifield1990} assuming a Plummer stellar profile. 
        Nevertheless, given the uncharacteristic way in which UDGs fill their dark-matter halos, we have also carried out a systematic check of how well various dispersion-based mass estimators are able to determine the masses of UDGs, which have half-light radii similar to $r_{\rm max}$ and a stellar component which contributes significantly to the central density. This comparison was motivated by the analysis of \cite{errani2018}, who show that estimators of the form $M(a~r_e)=b \sigma_{\rm los}^2 r_e/G$ \citep[e.g.][]{walker2009,wolf2010} are systematically biased depending on the shape of halo density profile and the stellar segregation $(r_e/r_{\rm max})$. In particular, we find that the above estimators underpredict the dispersions of UDGs (when compared with the line-of-sight Virial Theorem) by as much as $40\%$ (i.e.,~masses inferred from UDG dispersions may be overestimated by a factor of $2$). Regardless, we expect the dynamics of most UDGs to be more similar to field dwarfs than the Milky~Way.

        A somewhat counterintuitive prediction of this analysis is that the
        mass within the half-light radius is not significantly
        altered by the stripping process for systems in cored halos. 
        The galaxy expands as mass is lost in the center of the halo, keeping the mass
        within the half-light radius roughly constant (compared with the total amount of mass loss) and resulting in only a slight decrease in the predicted stellar velocity dispersion \citep{errani2015}.
        Overall, UDGs in our analysis manifest a
        $\sim0.07~{\rm dex}$ bias toward low $\sigma_{\rm los}$ at a given
        $M_*$. Apart from this bias, the relationship between $\sigma_{\rm los}$ and $M_*$
        for UDGs largely mirrors
        that of the field population, as dark-matter stripping occurs
        concurrently with stellar expansion (see Fig.~\ref{fig:dispersions}). There is a notable lack of UDGs with high stellar mass and high velocity dispersion in our model, as the highest mass systems requires the most tidal stripping (see Fig.~\ref{fig:magsize}). This is in contrast with predictions from feedback models,
		in
        which UDGs with the highest stellar mass formed in the most
        massive halos at late times, resulting in \emph{higher} velocity dispersions in the highest mass UDGs. Alternatively, if
        UDGs are formed in Milky~Way-like halos, one would expect a uniformly
        high stellar velocity dispersion, regardless of $M_*$, again contrary to
        our predictions. We find
                that $\sigma_{\rm los}$ decreases slightly at high $r_e$,
                in contrast with predictions from feedback models, which suggest a flat relationship between $r_e$ and $\sigma_{\rm los}$. Despite the dramatic tidal effects experienced by these
        systems, we predict that the baryonic tracers of host dark-matter halos
        should not significantly deviate from those of field dwarfs.

        This appears to be at odds with the large dispersion observed in
        DF~44 \citep{vandokkum2016}. While none
        of the UDGs in our statistical analysis are this dark-matter dominated,
        we would not expect them to be present if they were significant outliers
        compared with the overall galaxy population at infall given the size of
        our simulated sample. We find that the size, stellar mass, and velocity
        dispersion of DF~44 can be explained if the infalling galaxy started out
        in a more concentrated halo and with a larger than usual half-light
        radius. As a specific example, consider a mock galaxy with a stellar
        mass of $7.3\scinot{8}$~\msun{} and half-light radius of $3.7$~kpc
        that
        experiences $87\%$ mass loss within $r_{\rm max}$ over the course of
        two pericentric passages. Should such a galaxy live in an
        $8.5\scinot{10}$~\msun{} halo ($0.3\sigma$ above what our
        abundance-matching prescription predicts) with a concentration ($c$) of
        $18.3$ ($\sim2.7\sigma$ above the cosmological mass-concentration relation) 
        before it fell into Coma, the resulting
        halo would have a line-of-sight velocity dispersion of $42~{\rm km~s}^{-1}$ at $z=0$, 
        consistent with the measured
        dispersion of DF~44.\footnote{Placing DF~44 in a cored halo with an outer
        slope of $5$ (see Sec.~\ref{sec:massloss}) and
        $M_{\rm vir}=10^{10}~\msun{}$ is consistent with the observed
        stellar dispersion. In contrast, \cite{vandokkum2015b} place
        DF~44 in a $10^{12}~\msun{}$ NFW halo to explain the stellar velocity
        dispersion.} A similar scenario can explain the observed globular
                cluster dynamics of VCC~1287 and DFX1.

While the large sizes of these systems at infall lie well outside the predicted size-mass relation for quiescent systems ($3.7$~kpc is $>4\sigma$ away from the quiescent size-mass relation), it is not far beyond the star-forming size-mass relation.
In fact, this discrepancy ($r_e=3.7$~kpc is $1.4\sigma$ above the blue size mass relation for 
$M_*=3.5\scinot{8}~\msun{}$) may be lower if the intrinsic scatter in the 
                	size-mass relation is larger for dwarfs, as expected by some models as a consequence of star-formation feedback or dark-matter self-interactions
                	\citep{pontzen2014,chan2017,vogelsberger2014} and in line with the observed presence of such large systems in the field \citep{leisman2017}.
                	While our use of the size-mass relation for red galaxies was chosen to jointly match
                	the overall properties of the dwarf-elliptical and UDG populations in clusters (see 
                	   Sec.~\ref{sec:infallsizes}), the presence of UDG progenitors that more closely follow the size-mass 
                	   relation for star-forming systems before stripping may be necessary to produce the large 
                	   dispersions of these UDGs.
                       Overall, while a $M_*=3.5\scinot{8}$~\msun{} and $r_e=3.7$~kpc system initially seems to be an unlikely outlier in our model, such large systems should not be unexpected, and have indeed been observed in the field \citep{leisman2017}.
                      
        Similarly, as tidal stripping preferentially removes mass in the halo outskirts, our scenario allows
        	for the possibility of UDGs that appear baryon-dominated in their outskirts, like the recently identified UDG DF2 \cite{vandokkum2018}. In our scenario, if a dwarf with $M_*=2\scinot{8}$~\msun{} and $r_e=0.3$~kpc in a cored $M_{\rm vir}=5\scinot{10}$~\msun{}, $c=6.7$ halo lost $99.97\%$ of its mass, it would have $M_*=1.8\scinot{8}$~\msun{}, and $r_e=2.2$~kpc at $z=0$. The resulting dark-matter halo, would have a mass of only $9.1\scinot{7}~\msun{}$ and a low velocity dispersion consistent with the observations \citep{martin2018}.
        
        In summary, while systems like DF~2 and DF~44 should be rare among
        the UDG population in our scenario, the observed properties of these
        objects are consistent with unusual galaxies made more extreme through
        tidal stripping.

	\subsection{Tidal Features}
	A unique prediction of this formation scenario is that UDGs will be
        elongated along their orbital axes and exhibit distinct tidal features.
        Simulations of tidally-stripped systems find that axis ratios of
        galaxies in cored halos increase significantly after multiple
        pericentric passages \citep{errani2015}. Furthermore, S-shaped tidal
        features have been observed in satellites with recent tidal interactions
        \citep{choi2002,odenkirchen2003} as well as simulations of close
        host-satellite interactions \citep{johnston2002,johnston2008}.  This
        prediction is supported by the fact that UDGs in Coma tend to be aligned
        towards the cluster center \citep{yagi2016,burkert2017}. However,
        although tidal features have been observed around some UDGs
        \citep{toloba2016,muller2018}, \citet{mowla2017} find an absence
        of S-shaped tidal features out to $7$~kpc around UDGs in Coma. This
        absence of S-shaped tidal features establishes a lack of recent stellar
        stripping among most Coma UDGs. However, such features are not expected
        to persist long after a pericentric passage. Specifically,
        \cite{penarrubia2009} find that such features expand away from the
        satellite at roughly half the rate of the stellar velocity dispersion.
        Assuming that S-shaped tidal features expand away from UDGs at this
        rate, they would reach $7$~kpc from the UDG center $0.68$~Gyr after a
        pericentric passage. Among tidally-stripped UDG in Illustris-dark, $13\%$ 
        experienced a pericentric passage in the
        past $0.68$~Gyr, suggesting that S-shaped tidal features should be
        observable in this fraction of UDGs in Coma, marginally consistent with
        the observations of \citet{mowla2017}. Furthermore, the S-shaped
        features searched for by \citet{mowla2017} are expected to become
        aligned along the orbital direction quickly \citep{penarrubia2009},
        significantly reducing the predicted distortion of UDG morphologies.

	\subsection{Radial Distribution}
	The spatial distribution of UDGs within clusters can provide a valuable
        piece of information regarding UDG formation and
        evolution. Among Illustris subhalos, those hosting tidally-stripped UDGs largely
        mirror the subhalo distribution, with a slight ($\sim4\%$) in the cluster centers ($0.1\times R_{200}$). 
        Given that $53\%$ of tidally-stripped
        UDGs in cored halos experience two or fewer pericentric passages, it
        makes sense that tidally-stripped UDGs are not exclusively found in
        cluster centers.
        
	On the other hand, our finding that UDGs tend to be slightly
        overabundant in cluster centers relative to subhalos of similar mass is
        inconsistent with the observation of a deficit of UDGs in cluster
        centers \citep{vandokkum2015,vanderburg2016}. This may point to a
        breakdown of our model in the most extremely disrupted systems. Our
        analysis produces a number of remarkably tidally-stripped ($>99\%$ mass
        loss) systems in the central regions of the clusters. Our model for tidal stripping is not as well constrained for systems with this degree of mass loss, such that UDGs with a such a high degree of stripping may loose mass more quickly than our model predicts. For example, the assumption that tidal stripping only occurs at pericenter may break down in the central regions as tidal effects become more important throughout the orbit. In this case, we may underestimating how many UDGs in cluster centers are stripped to the point that they are below our detection threshold or completely destroyed. 
	
		\begin{figure}
			\centering
			\includegraphics[width=1\linewidth]{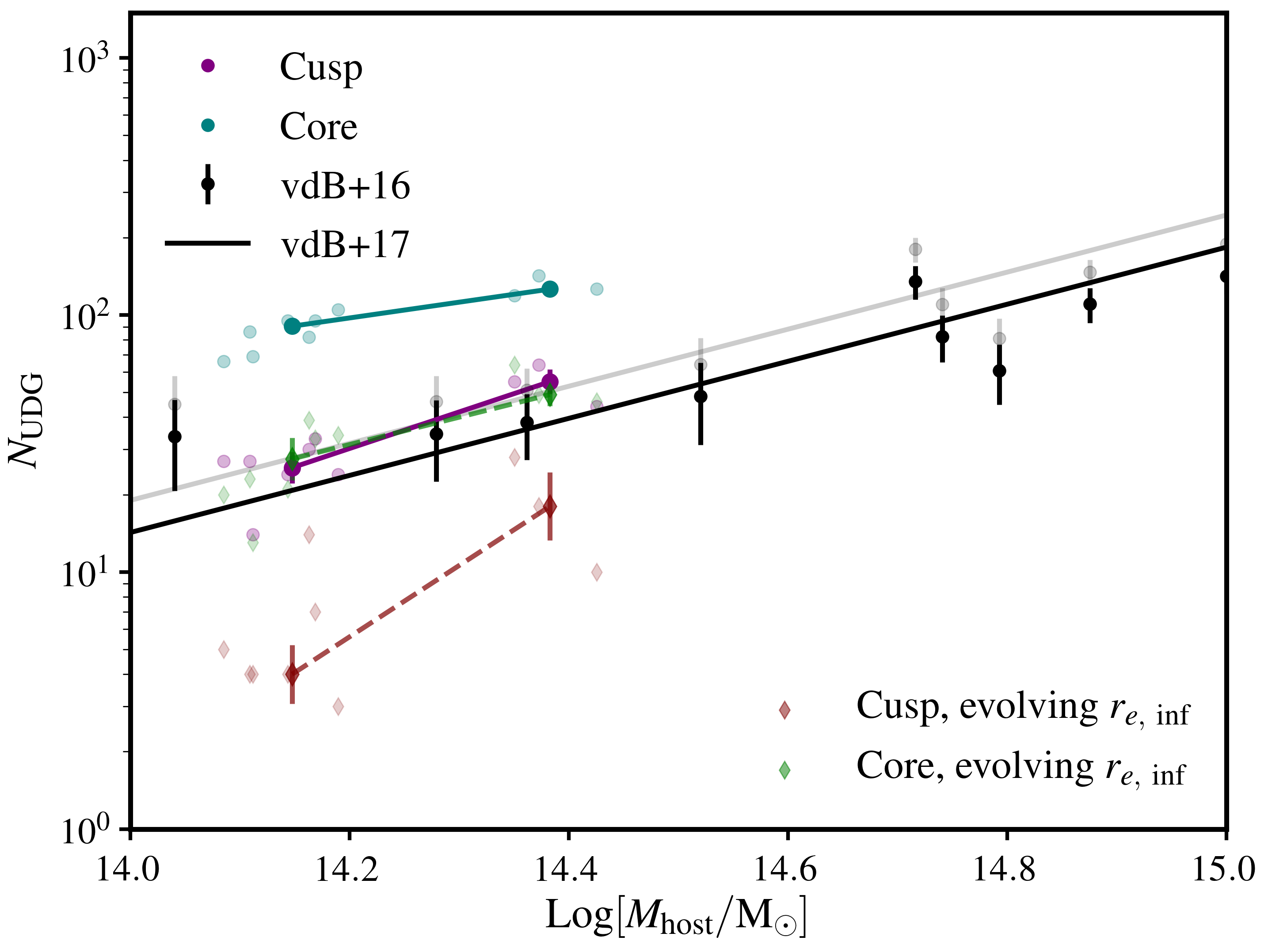}
			\caption{The abundance of tidally-stripped UDGs as a function of host
				halo mass for UDGs hosted in cuspy (teal) and cored (purple)
				dark-matter halos. For comparison, the black points and back line
				correspond to observed UDG abundances from \protect
				\cite{vanderburg2016} and the fit to the UDG abundance as a function
				of halo mass from \protect \cite{vanderburg2017}, respectively.
				For our fiducial model, the abundance of cored UDGs is a factor of $4$ greater
				than the observed UDG abundance and the abundance of cuspy UDGs matches the observed UDG abundance. However, if the starting sizes evolve with redshift according to $(1+z)^{-0.6}$, the abundance of cored UDGs (as shown by the light teal diamonds) agrees with the observed abundance within and the abundance of cuspy UDGs (as shown by the purple diamonds) significantly underestimates the UDG abundance.}
			\label{fig:nudgvshost}
		\end{figure}

        \subsection{Abundance}
        \label{sec:udgabundance}

        The increase in UDG abundance with increasing cluster mass
        \citep{vanderburg2016,vanderburg2017} provides a powerful constraint on
        the strength of environmental processes in shaping the UDG population.
        For instance, \cite{vanderburg2017} find that UDG abundance increases
        with increasing host mass as
        $N=19\times (M_{200}/10^{14}~\msun{})^{1.11}$, motivating an
        environmental-dependent formation mechanism. In \cite{vanderburg2017},
        the UDG abundance is determined by subtracting the abundance of UDGs
        surrounding clusters by the UDG abundance measured using the same
        procedure on a random region of the sky. However, this subtraction still
        overestimates the abundance of cluster satellites by counting objects
        infalling onto the cluster. To estimate the magnitude of this effect, we
        measure the ratio between the number of halos that would be counted as
        cluster members in the \cite{vanderburg2017} sample and the true
        satellite abundance in the Illustris-dark simulation. In \cite{vanderburg2017}, satellites are
        identified as galaxies within the projected $R_{200}$ of the cluster and
        with a line-of-sight velocity within $5\sigma$ of the cluster, where
        $\sigma$ is the velocity dispersion of the cluster from
        \cite{sifon2015}. We find that the \cite{vanderburg2017} abundance
        measurements overestimate the true abundance by a factor of $1.3$, with
        no significant dependence on either subhalo mass or cluster mass. Thus,
        we take the measured UDG abundance divided by this factor of $1.3$ as
        the true UDG abundance within $R_{200}$.
        
	Figure~\ref{fig:nudgvshost} shows the relationship between the abundance
        of tidally-stripped UDGs and host mass found within our analysis. As
        expected, the abundance of tidally-stripped UDGs increases with
        increasing halo mass; we find
        $N=67\times \left(M_{200}/10^{14}~\msun\right)^{0.76}$ for cored halos and $N=18\times \left(M_{200}/10^{14}~\msun\right)^{1.24}$ for cuspy halos. This model overproduces the UDG abundance by a factor of $3.8$ for cored halos and $41\%$ for cuspy halos.
        However, if the starting size-mass relation evolves weakly with redshift, following $(1+z)^{-0.6}$, the predicted abundance
        matches the observed for cored halos and the UDG size distribution is relatively unchanged (see Appendix~\ref{sec:appendix2}). Furthermore, the final size-mass relation for cored subhalos in this case is actually more consistent with the observed
        size distribution of dwarfs ($5\scinot{7}<M_*/M_\odot<2\scinot{8}$) in the Fornax Cluster \citep{munoz2015} than with the fiducial size-mass relation (KS-test $p$ value of $0.4$ compared with $0.002$). 
        
        Among clusters in our sample, the UDG abundance is a relatively constant fraction of the total subhalo abundance for both cored and cuspy subhalos,
        in contrast with the observed super-linear increase in UDG abundance with halo mass.
        However, the comparatively narrow range in host halo mass probed by this analysis limits our ability to compare this trend with the observed relation. If we make the assumption that the UDG fraction depends on the host mass within $200$~kpc (the mean pericenter distance for among our sample), it is possible to extrapolate the $N_{\rm UDG}-M_{\rm host}$ relation. Within our sample, we find:
        
		\begin{equation}
		\label{eqn:udgfrac}
		{\rm Log}\left<\frac{N_{\rm UDG}}{N_{\rm sub>10^{9.8}~M_\odot}}\right>=0.76~ {\rm Log} \left[\frac{M_{\rm host}(200~{\rm kpc})}{10^{13}~\msun{}}\right]-1.39\; ,
		\end{equation}
		where ${N_{\rm UDG}}/{N_{\rm sub>10^{9.8}~M_\odot}}$ is the fraction of subhalos with infall masses $>10^{9.8}$~\msun{} that become UDGs and $M_{\rm host}(200~{\rm kpc})$ is the host mass within $200$~kpc.
		Extrapolating this relation down the hosts of $10^{12}$~\msun{}, assuming the subhalo mass function of \cite{jiang2016} and the \cite{prada2012} mass-concentration relation, this result suggests
		$N_{\rm UDG}\propto M_{\rm host}^{1.2}$, roughly consistent with the observed trend in this mass range. Ultimately, although our fiducial model does not match the observed UDG abundance exactly, it is roughly consistent with observations.
	
		Notably, as this method of UDG formation requires subhalos to orbit a massive cluster for many Gyr, it suggests that the relative UDG abundance is lower among recently-formed clusters. This suggests that more relaxed clusters and clusters with higher concentration values should have a higher UDG abundance than less relaxed clusters and clusters with lower concentration values of similar mass. Additionally, this suggests that the relative abundance of UDGs decreases with increasing $z$ -- we find that the UDG abundance has decreased by $83\%$ by $z=0.5$ (whereas the cluster subhalo abundance has decreased by only $24\%$).
	
	\section{Conclusions}
        \label{sec:conclusions}

	In this work, we have shown that tidal stripping of dwarf galaxies
        within clusters is able to reproduce the observed properties and
        abundance of UDGs remarkably well. The principal results of our analysis
        are as follows:

        \begin{itemize}[leftmargin=0.25cm]
        
        \item In our modeling of tidal stripping, we find that \emph{cuspy}
          halo do not experience the required mass loss to generate a
          substantial number of large cluster UDGs. In contrast, tidal forces acting
          upon dwarf galaxies hosted by $\sim10^{10}-10^{11}~\msun{}$
          \emph{cored} halos do produce a cluster UDG population with properties
          (e.g.~stellar mass, size, metallicity, etc.) that broadly agree with
          observed UDG samples. \\ 

        \item The distribution of observed half-light radii for local UDGs is
          reproduced by that of tidally-stripped UDGs in cored dark matter
          halos. In particular, the most extended UDGs with $r_{e}>3$~kpc, which
          observations estimate to be $\sim11-17\%$ of the cluster UDG
          population, comprise $5\%$ of our tidally-stripped, simulated
          sample. \\

        \item The abundance of tidally-stripped UDGs in cored halos increases
          with increasing cluster mass according to
          $N \propto M_{\rm host}^{0.76}$, roughly consistent with the observed abundance.\\

        \item A number of concrete predictions arise from our model for
          UDG formation:
          
          \itemize
        \item 
        While small UDGs have a range of stellar ages, the largest
          tidally-stripped UDGs ($r_e>3$~kpc) should host significantly older stellar populations,
          with $89\%$ of such systems $>4$~Gyr
          old.\\

        \item Dark matter halos hosting UDGs have similar masses to typical
          dwarfs of the same mass. They should be more centrally concentrated
          than dwarfs at the same mass because dark matter is preferentially lost in the halo outskirts.\\
          
        \item At fixed host mass, UDG abundance should be correlated with cluster age/concentration and inversely correlated with redshift.
    	\end{itemize}
        
        The main uncertainty of this model is the initial size-distribution of dwarf galaxies before they experience tidal stripping and heating -- the difference between the star-forming and quiescent size-mass relation can be as much $0.2$~dex within our mass range. Compounding this difference with any possible redshift evolution and the different amounts of intrinsic scatter, this uncertainty easily has the largest effect on our analysis, and we advocate for further observational and theoretical work into investigating the processes processes involved in transforming large dwarf-irregular galaxies to smaller dwarf-ellipticals. Beyond that, the treatment of tidal stripping and heating among systems with baryon-dominated centers as well as systems with extreme mass loss and needs significant further investigation. In particular, the population of extremely large UDGs (i.e. those underproduced by our model) is composed of systems with $>99\%$ mass loss -- beyond the bounds of the \cite{errani2018} tracks.
 
        We have presented a scenario for the formation of ultra-diffuse galaxies
        in clusters motivated by the presence of cores in field dwarf
        galaxies. We have shown that this scenario provides a compelling
        explanation for the abundance and radial alignment of ultra-diffuse
        galaxies and it makes concrete predictions that can be tested in the
        near future.

\section*{Acknowledgments}

The authors are grateful to Remco van der Burg for helpful discussions related
to the observed abundance of UDGs. Additionally, TMC would like to thank Andrew
Graus for many helpful discussions regarding $N$-body simulations and Sheldon Campbell for his help
with the analysis. Lastly, we are grateful to the anonymous reviewer, whose suggestions greatly improved this paper. This work was
supported in part by NSF grants AST-1518257 and PHY-1620638. Support for this work was provided
by NASA through grants (AR-13242 and AR-14289) from the Space Telescope Science
Institute, which is operated by the Association of Universities for Research in
Astronomy, Inc., under NASA contract NAS 5-26555.

This research made use of {\texttt{Astropy}}, a community-developed core Python
package for Astronomy \citep{astropy}. Additionally, the Python packages
{\texttt{NumPy}} \citep{numpy}, {\texttt{iPython}} \citep{ipython},
{\texttt{SciPy}} \citep{scipy}, and {\texttt{matplotlib}} \citep{matplotlib}
were utilized for the majority of our data analysis and presentation.

\bibliography{udgpaper}

\appendix
\section{Considering the Effects of Baryons on Cored Dark-Matter Halos}
\label{sec:appendix}
In this Appendix, we discuss in more detail how our analysis is affected by the presence of baryons in cored halos, particularly 
when 
considering 
baryon-dominated systems (see Sec.~\ref{sec:massloss}).

\begin{figure}
	\centering
	\includegraphics[width=1\linewidth]{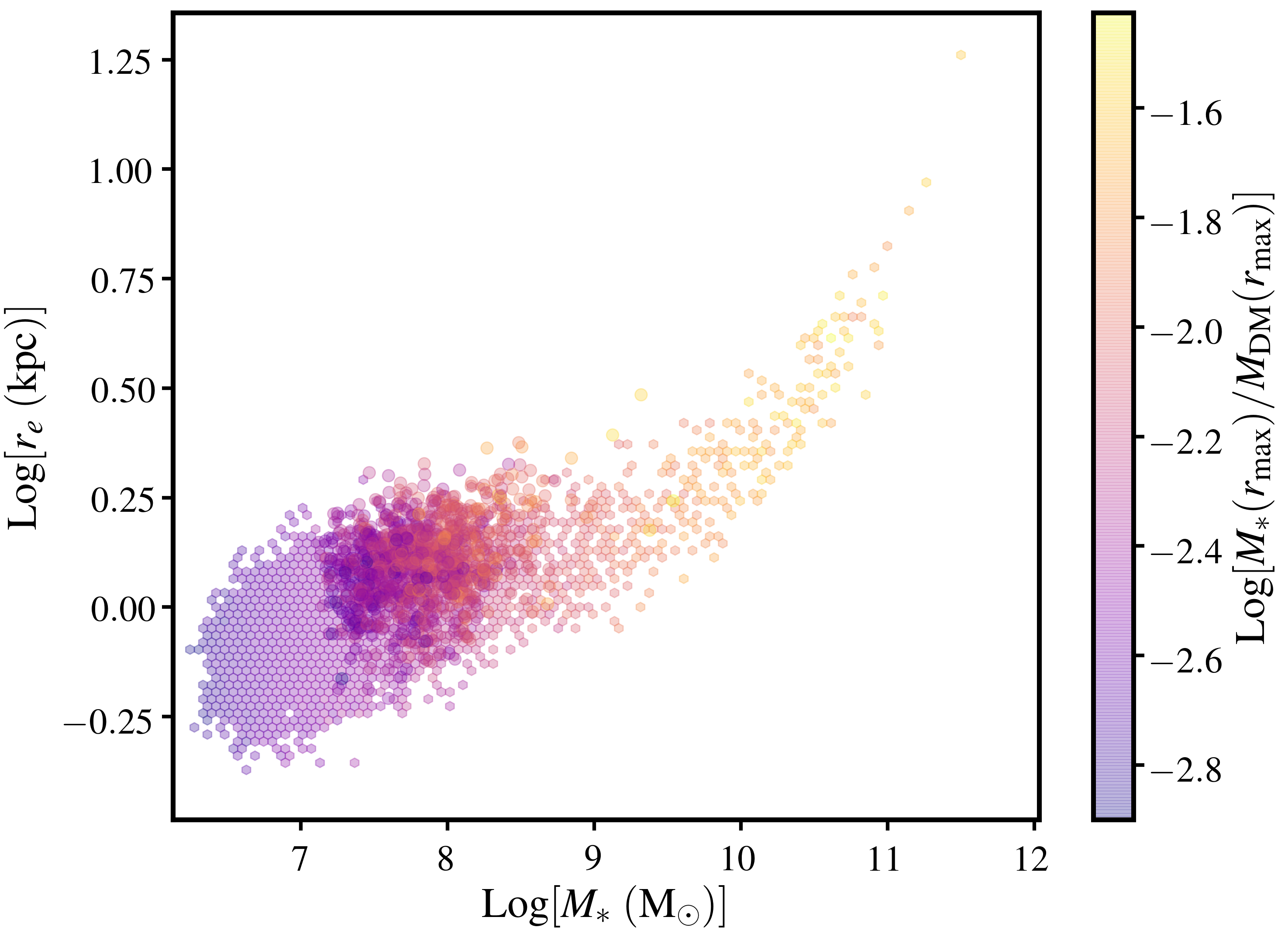}
	\caption{The size-stellar mass relation at infall for our simulated systems in Illustris-dark, color-coded by the stellar 
		to dark matter mass ratio within $r_{\rm max}$, with systems that evolve into UDGs highlighted. For all 
		systems, the baryonic mass represents a very small fraction of the dark-matter mass.}
	\label{fig:mrmaxratio}
\end{figure}

Our assumption that baryon-dominated systems undergo the same dark-matter stripping as dark matter-dominated systems is motivated by the fact that the inclusion of baryonic mass has a negligible impact on the $V_{\rm max}$ and $r_{\rm max}$ of a halo (see Fig.~\ref{fig:mrmaxratio}), although they could dominate within the half-light radius for some progenitors.
Nevertheless, the presence of baryons may cause the halo to be more concentrated.
Observations are consistent with a  
range of core sizes spanning from the NFW scale radius to the galaxy's half-light radius 
\citep{oman2015,oh2015}, and this wide variation 
has been observed in models including SIDM and baryons 
\citep{vogelsberger2014,kamada2017,creasey2017} as well as baryonic feedback \citep{fitts2017,dicintio2014}.
Our cored model, taken to match  models including SIDM and baryons for galaxies in our sample 
\citep{kaplinghat2016,kamada2017,elbert2018},  produces cores that extend to the scale radius of the halo.
Nevertheless, we test how varying the 
core size affects our results by rerunning our analysis with the halos at infall generated by following the 
prescription of \cite{dicintio2014b}, which is designed to fit halos affected by baryonic feedback with a 
mass-dependent profile shape. For the $r_e$ and $M_*$ tracks, we take $\gamma$ to be the larger of $\gamma$ 
determined from $\delta_{1/2}$ or $\gamma$ from \cite{dicintio2014b}.
\begin{figure}
	\centering
	\includegraphics[width=1\linewidth]{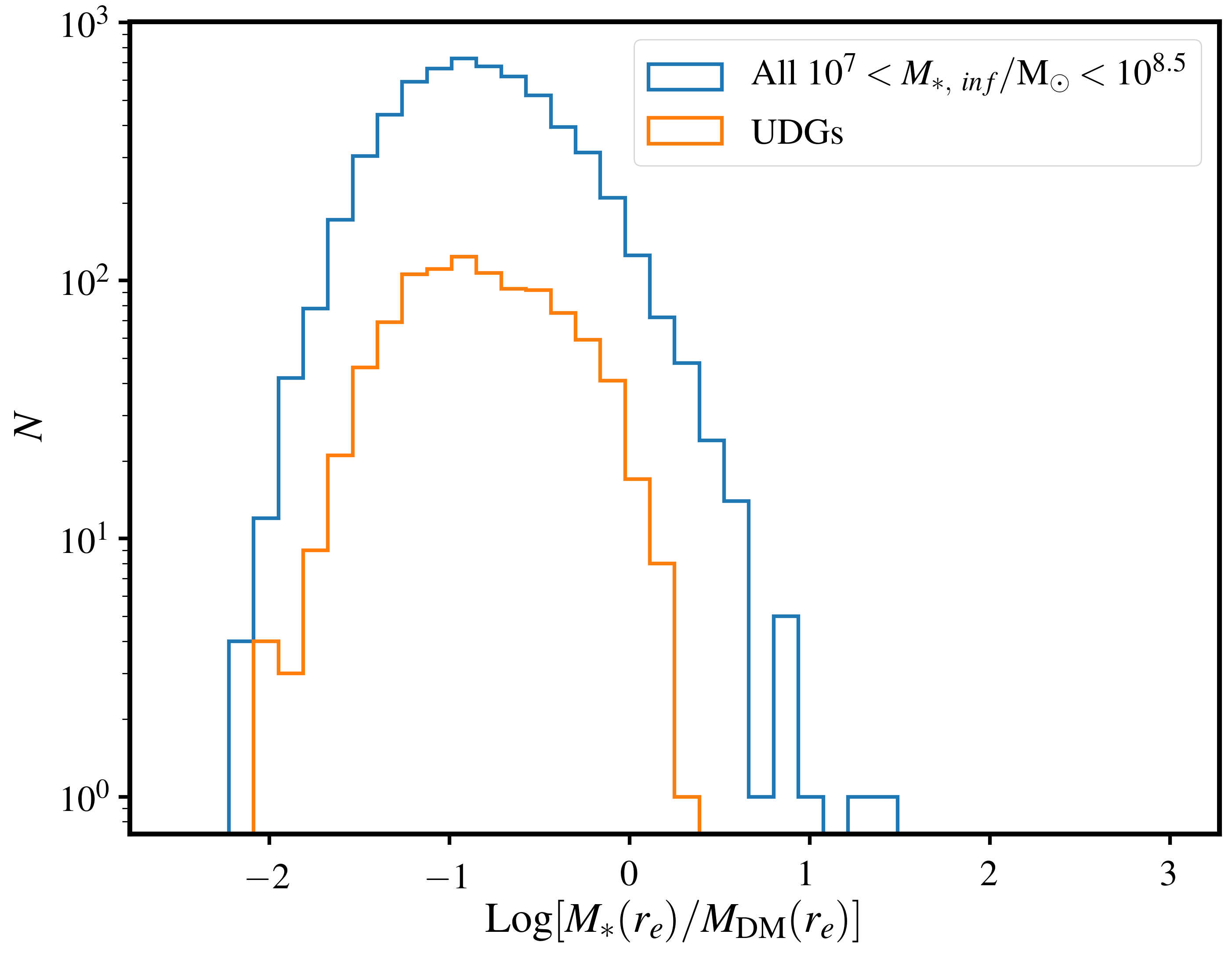}
	\caption{The distribution of $\delta_{1/2}$ for all simulated systems in Illustris-dark within the mass range ($10^7 < 
		M_*/M_\odot < 10^{8.5}$, blue line) and for just those systems the become UDGs (orange line).  Overall, 
		baryon-dominated systems are a negligible component of the starting population, and the exact treatment 
		of 
		them does not significantly change the conclusions of our analysis.}
	\label{fig:deltahist}
\end{figure}
As these halos have a shallower outer slope than our fiducial model, they experience slightly more mass loss overall. However, as they are more cuspy than our fiducial model, they experience less size growth. The resulting abundance increases by less than $1\%$ compared with the fiducial abundance, but the size distribution is slightly steeper than observed at $n(r)\propto r^{-3.5}$. However, if we force the halos to have an outer slope of $3$ at infall, the UDG abundance decreases by $12\%$ and the size distribution steepens to $n(r)\propto r^{-4.3}$.
Alternatively, many authors suggest that a more isothermal core is 
appropriate 
\citep[e.g.~][]{oh2015}, which 
transitions more sharply from the outer slope of $3$ to a central core than Eqn.~\ref{eqn:corenfw}. If our cored 
systems are modeled as a psudo-isothermal profile:
\begin{equation}
\label{eqn:iso}
\rho(r)=\frac{\rho_s}{ \left(1+\left(\frac{r}{r_s}\right)^2\right)^{1.5}} \; ,
\end{equation}
matched to the $V_{\rm max}$ and $r_{\rm max}$ of the subhalo at infall, we observe a similar effect as the \cite{dicintio2014b} halos. More mass in the halo outskirts results in increased stripping, but increased $\gamma$ values (because of the lower core density) results in fewer large UDGs. Additionally, the lower $M_{\rm vir,~inf}$ values of these halos (at fixed $V_{\rm max}$ and $r_{\rm max}$), results in a slightly lower UDG abundance. The abundance decreases by $26\%$ and the size distribution steepens to $n\propto r^{-4.1}$.
Altogether, the halo occupation, stellar population, and environmental dependence predicted by our model are 
independent of the exact implementation of the cored profile, and even in extreme cases, the abundance 
produced by this analysis predicts a substantial UDG population.

\begin{figure}
	\centering
	\includegraphics[width=1\linewidth]{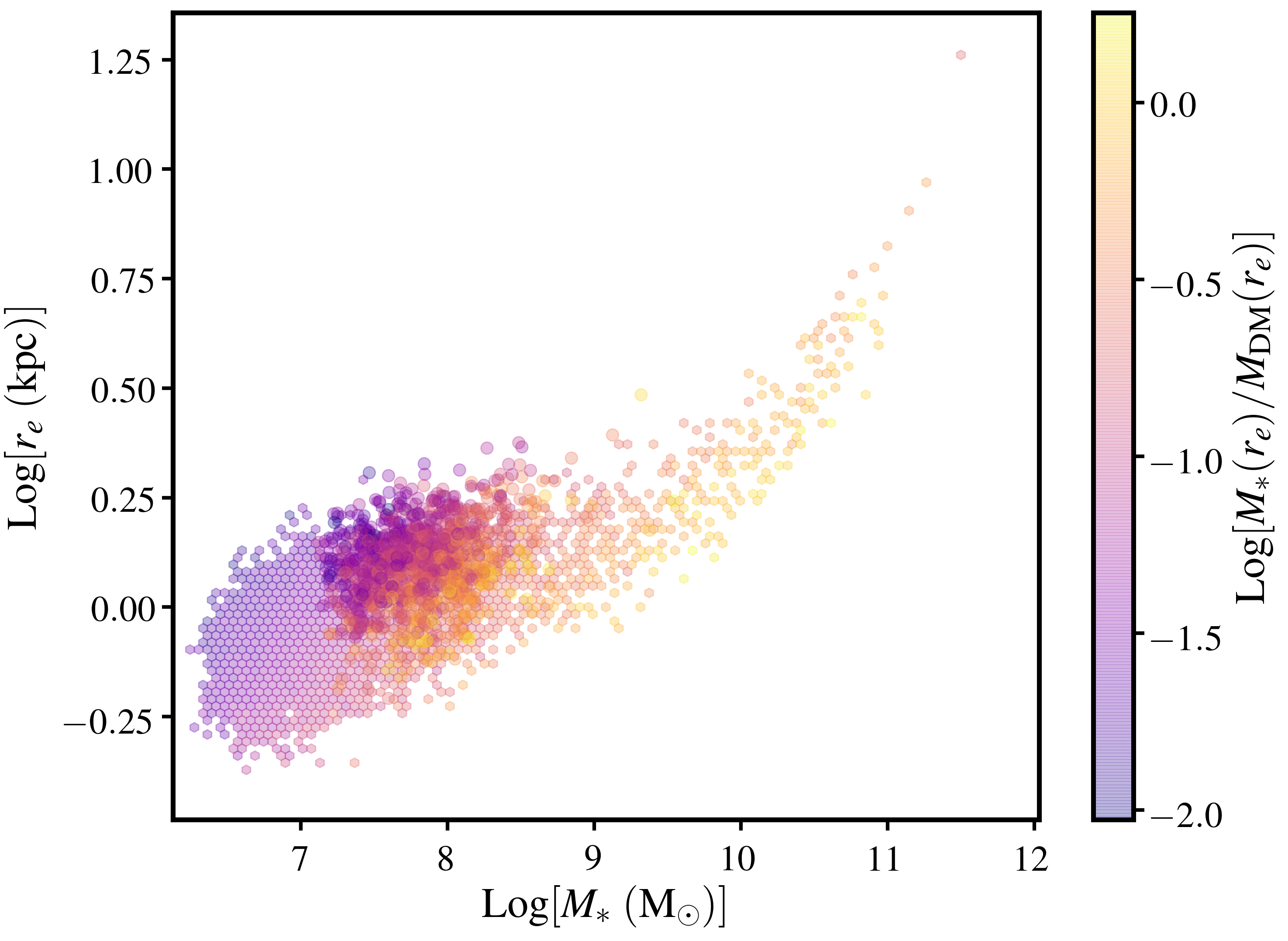}
	\caption{The infall size-mass relation, color coded by the stellar mass within $r_e$, with systems that end up 
	as 
		UDGs highlighted. Although systems with $\delta_{1/2}$ as high as $3$ can become UDGs, the vast 
		majority of 
		systems that become UDGs are mildly dark-matter dominated. }
	\label{fig:mr12ratio}
\end{figure}

Our prescription for dealing with tidal heating in baryon-dominated systems is designed to reproduce 
the results from \cite{errani2018} for dark matter-dominated systems and limit the effects of tidal heating in 
baryon-dominated systems. Overall, this prescription has the desired effect: although some baryon-dominated 
systems end up as UDGs, the majority are 
dark-matter dominated (only 
$5\%$ of systems within 
$10^7<M_{*,~{\rm inf}}/M_\odot<10^{8.5}$ have $\delta_{1/2}$ values greater than $1$, and $36\%$ have 
$\delta_{1/2}$ 
values less than $0.1$, see Fig.~\ref{fig:deltahist},~\ref{fig:mr12ratio}).
While the 
adoption of an intermediate $\gamma$ is an important consideration for our full model, tidal heating of only dark matter-dominated systems in cored profiles is able to produce a reproduce many of 
the observed UDG properties.
Although our analysis is sensitive to the value of 
$\delta_{1/2}$ for which $\gamma$ begins to change, our results do not change significantly
if 
systems with $\delta_{1/2}$ above $0.5$ are assigned higher $\delta_{1/2}$ values. 
Similarly, as less than $4\%$ of systems have $\delta_{1/2}$ between $1$ and $2$, our analysis is not sensitive to 
exactly how $\gamma$ depends on $\delta_{1/2}$ within that range (see Fig.~\ref{fig:deltahist}).
Lastly, given evidence that 
dwarf galaxies of all morphologies have flat stellar profiles 
\citep{faber1983,vandokkum2015,yagi2016}, the inner slope of the baryonic component is likely $0$, so 
$\gamma=0$ may be appropriate even in 
baryon-dominated cases. 

\begin{figure}
	\centering
	\includegraphics[width=1\linewidth]{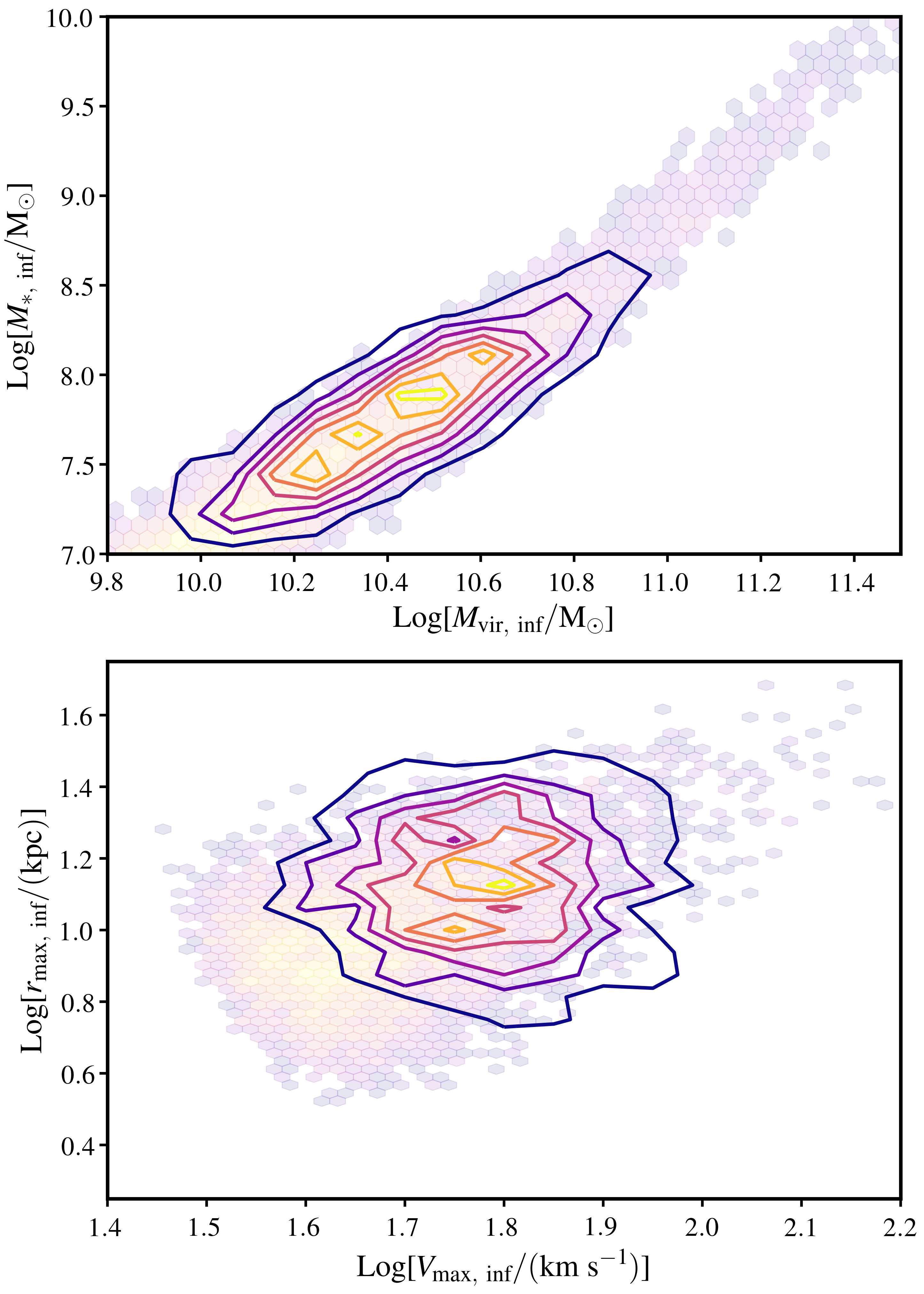}
	\caption{The infall abundance matching relation (top) and $V_{\rm max}-r_{\rm max}$ relation (bottom) 
	color-coded by relative abundance in Illustris-dark, with systems that end up as 
		UDGs identified as contours. In our scenario, halos hosting UDGs are cosmologically representative at infall.}
	\label{fig:amrelation}
\end{figure}

Finally, we emphasize that the dark-matter halos hosting UDGs are not drawn from a particularly biased distribution. 
Figure~\ref{fig:amrelation} show both the stellar-mass-halo-mass relation and $V_{\rm 
max}-r_{\rm max}$ relations of field halos in Illustris-dark, with systems that become UDGs highlighted. Although the 
systems that 
become UDGs tend to be larger at infall (Fig.~\ref{fig:mrmaxratio},~\ref{fig:mr12ratio}), the halos hosting these 
systems are 
not particularly special.

\section{UDG Properties Assuming an Evolving Size-Mass Relation}
\label{sec:appendix2}
\begin{figure}
	\centering
	\includegraphics[width=1\linewidth]{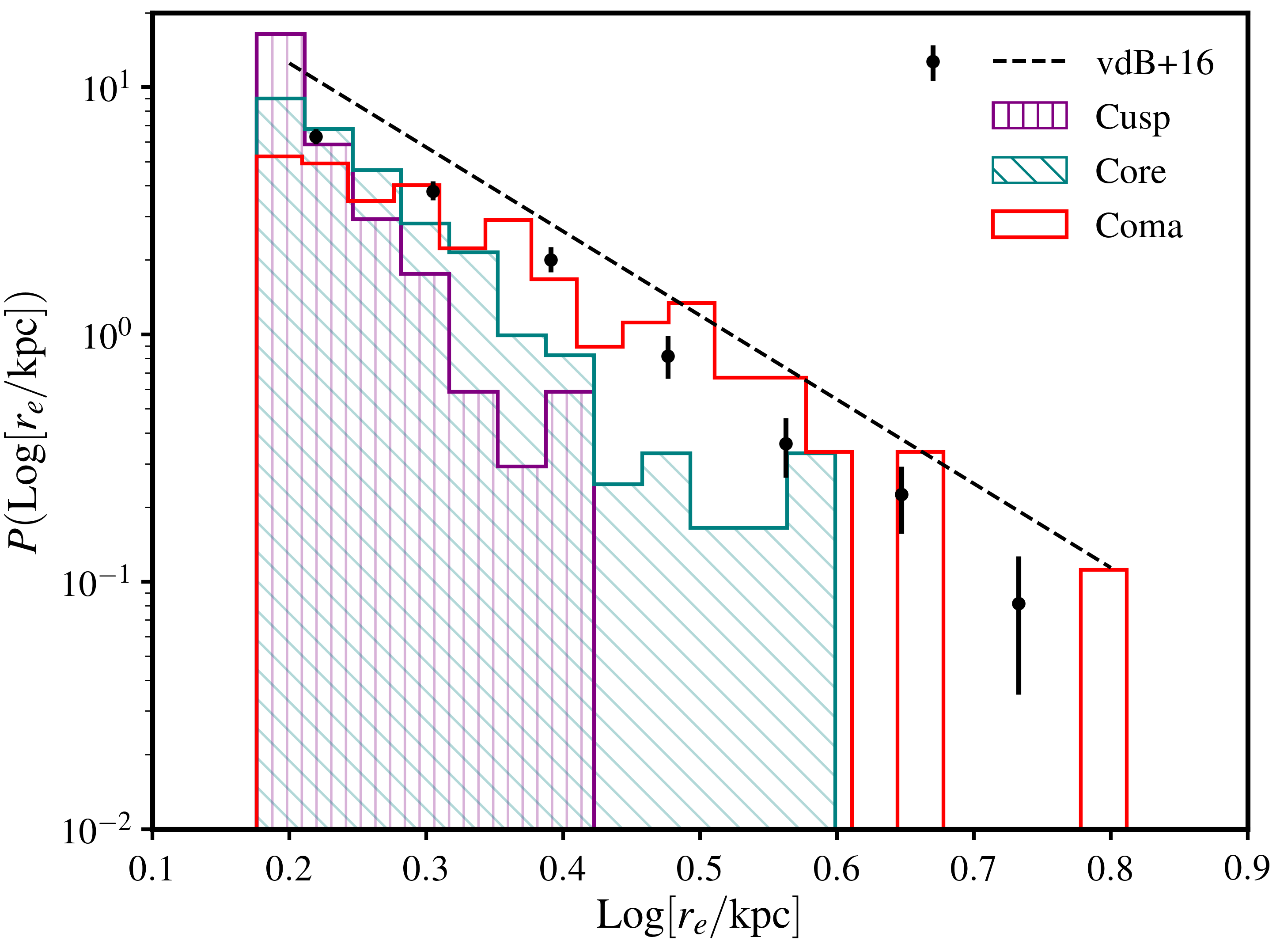}
	\caption{The size-distribution of UDGs produced through tidal stripping and heating, assuming a weakly evolving size-mass relation. Although this weak evolution decreases the abundance of the largest UDGs, the distribution is largely unchanged.}
	\label{fig:sizedistevolve}
\end{figure}

To reproduce the observed UDG abundance, we postulate a size-mass relation that evolves weakly with redshift, following $(1+z)^{-0.6}$. Here, we discuss the properties of UDGs produced in this scenario. First, although the UDG size distribution is slightly steeper than our fiducial model (see Fig.~\ref{fig:sizedistevolve}), it still produces a large population of large UDGs, with $2\%$ of UDGs greater than $3$~kpc. Similarly, we find that the mass distribution and velocity dispersions are not significantly changed from the non-evolving case. Lastly, while the evolving case results in younger UDGs overall, the trend between UDG age and size persists.

\end{document}